\documentclass[prb,aps,showpacs,floatfix,groupedaddress,amsmath,twocolumn]{revtex4}

\usepackage[dvips]{graphicx}
\usepackage{dcolumn}
\usepackage{bm}
\usepackage{epsfig}
\usepackage{amssymb}

\begin{document}


\title{Correlation effects in quasi one dimensional electron wires}

\author{Luke Shulenburger,$^{1}$ Michele Casula,$^{1}$ Gaetano
Senatore,$^{2,3}$ and Richard M. Martin$^{1}$}

\affiliation{ $^1$ Department of Physics, University of Illinois at
Urbana-Champaign, 1110 W. Green St, Urbana, IL 61801, USA\\
$^2$ INFM Democritos National Simulation Center, Trieste, Italy \\
$^3$ Dipartimento di Fisica Teorica dell' Universit\`a di Trieste, Strada
Costiera 11, 34014 Trieste, Italy }

\date{\today}

\begin{abstract}
We explore the role of electron correlation in quasi one dimensional quantum
wires as the range of the interaction potential is changed and their
thickness is varied by performing exact quantum Monte Carlo simulations at
various electronic densities. In the case of unscreened interactions with a
long range $1/x$ tail there is a crossover from
a liquid to a quasi Wigner crystal state as the density decreases. 
When this interaction is
screened, quasi long range order is prevented from forming, although a
significant correlation with $4 k_F$ periodicity is still present at low
densities.  At even lower electron concentration, exchange is suppressed and 
the electrons behave like spinless fermions.   Finally, we study the effect of electron
correlations in the double quantum wire experiment [Steinberg \emph{et
al.}, Phys. Rev. B \textbf{77}, 113307 (2006)], by introducing an accurate
model for the screening in the experiment and explicitly including the
finite length of the system in our simulations.  We find that decreasing the
electron density drives the system from a liquid to a state with quite
strong $4 k_F$ correlations. This crossover takes place around 
$22 ~ \mu m^{-1}$, near the density where the electron localization 
occurs in the experiment. The charge and spin velocities are also in
good agreement with the experimental findings in the proximity of the crossover.
We argue that correlation effects play an important role at the onset of the
localization transition. 
\end{abstract}


\maketitle

\section{Introduction}

It is well known that the effect of interactions in quasi one dimensional
(Q1D) systems of electrons, usually called ``quantum wires'', is enhanced
compared to higher dimensional systems.
There are universal properties described by the Luttinger liquid paradigm,
the \emph{effective} low energy theory which applies 
for strictly 1D models,\cite{luttinger,tomonaga,voit,chang,giamarchi-book}
such as spin-charge separation,charge localization, and conductance quantization.
However, the  microscopic details, such as the width and type of the transverse confinement or
the distance and shape of neighboring screening media, can have a large impact on the properties of
the Q1D systems, as they are very sensitive to the effective interaction.   These systems can be
realized in semiconductor structures, where there are elegant experimental 
studies,\cite{auslaender-science0,
soft-confinement-prl,soft-confinement-prb,auslaender-solid-state-comm,
auslaender-science1,yacoby,steinberg-tunneling-depleted-top-wire}
and it is essential to describe the system accurately for a realistic comparison of theory and experiment.

In this paper we study how the thickness, finite size, and screening affect
the phase boundaries of some universal features with a particular emphasis
on the charge localization and spin properties.  We address the issue of how
the electron correlation depends upon the microscopic details parametrized
in the interaction using ground 
state quantum Monte Carlo (QMC) methods\cite{RevModPhys_foulkes}
such as diffusion Monte Carlo\cite{reynolds_dmc,umrigar_dmc}, and its
lattice regularized version\cite{casula_lrdmc} which are ideal
numerical tools to study Q1D systems, since they provide exact results in
one dimension. Previous QMC studies regarded the determination of the LL
parameters for a Q1D system with screened
interactions,\cite{hausler-qmc-quantum-wires} and the ground state
properties of a model with a long-range Coulomb potential.\cite{casula-1d}
Here we compare various model interactions in a unified picture, with the
final goal of quantifying the role of correlation in the localization
transition found by Steinberg \emph{et al.}.\cite{steinberg-tunneling-depleted-top-wire}

With this aim it is particularly important 
to include both the effects of the long-range Coulomb potential
and the consequences of its screening in the interactions.  
In dimensions larger than one, the $1/x$ tail of the Coulomb pairwise potential
leads to a Wigner crystal phase of the homogeneous gas at low densities,
when the potential energy dominates over the kinetic
contribution.\cite{wigner-crystal,pines-book,ceperley-alder}  In this regime
the spin exchange drops to an exponentially small
value\cite{ceperley-exchange} as the overlap between unlike spin particles
is exponentially suppressed by the localization of the electrons. Therefore,
one of the signatures of the liquid-to-crystal transition is a decrease in
the spin stiffness. However, it is possible that other spin and charge phases
could exist in between.

In 1D the situation is radically different. It is well known that a
Luttinger liquid (LL) with $1/x$ interactions exhibits slowly decaying
charge-charge correlations, but no true long range order, as the quantum
fluctuations are stronger in lower dimensions.\cite{giamarchi-book}
Nonetheless there should be a crossover from a high density liquid to a low
density regime with quasi long-range charge order also called a
``fluctuating Wigner crystal''.\cite{schulz} However the LL theory does not
predict where the crossover happens, as the correlation function parameters
are not universal but depend on the details of the interaction.  Also the
interplay between charge and spin is quite unclear. Indeed the LL parameters
depend on the effective one dimensional potential in a non trivial
way.  For instance, the spin properties are strongly affected by its
short-range behavior which includes the effect of the thickness, as the
transverse dimension can effectively tune the spin exchange. On the other
hand, the quasi order of the charge degrees of freedom is stabilized by the
long range tail. The relative importance of the short versus long range
correlations is set by the microscopic model of the system.
Recently Fogler\cite{fogler-coulomb-tonks-prl,fogler-coulomb-tonks-prb}
proposed that a correlated state with very small spin exchange exists for
ultrathin wires at densities \emph{between} the liquid and the quasi Wigner
crystal phase. In the limit where the short-range part can 
be effectively described by an infinite repulsive contact interaction, 
this state can be related to a \emph{noninteracting} spinless Fermi system,
as in the Tonks-Girardeau gas.\cite{girardeau}
In a one dimensional system of fermions, the coexistence of strong
short-range repulsions and very long-range interactions leads to a peculiar
state, which Fogler termed a Coulomb-Tonks gas.

In previous theoretical work, quantitatively accurate studies of the
liquid-to-crystal one dimensional crossover have been carried out only for
inhomogeneous systems, with longitudinal extension controlled by an external
confinement, and where the finite (and very small) size allows one to solve the
problem by means of exact diagonalization.\cite{lanczos1,lanczos2,lanczos3}
However, the broken translational symmetry leads to quite different
properties, particularly in the charge and spin density profiles. 

From the experimental side, technological advances in the preparation of
cleaved edge overgrowth samples have enabled tunneling measurements between
two high mobility parallel wires, which probe striking features like the
spin-charge separations in the excitation spectra.\cite{auslaender-science1}
In a recent extension of the tunneling experiments, Steinberg \emph{et
al.}\cite{steinberg-tunneling-depleted-top-wire} applied a gate to the upper
wire in order to tune its electron density by charge depletion. Below a
critical threshold, measurements revealed a dramatic transition which can be
interpreted as the onset of localization in the wire.  Although 
it is believed that the transition is mainly driven by electron-electron
interaction effects as the liquid phase is in a ballistic regime, so far there is no
agreement between the critical density predicted by theory and the actual
experimental value.  Some features of the experiment, like the fringes of
the differential conductance in the liquid phase and the first two peaks of
the tunneling current in the localized phase, have been explained in an
independent particle picture\cite{soft-confinement-prb,soft-confinement-prl}
and at the mean field level.\cite{mueller}  The LL theory has been applied
to describe the general features of the tunneling current in the Wigner
state by assuming a spin incoherent regime.\cite{fiete} However it is
unclear at which density the spin degrees of freedom become incoherent, as
the spin velocity measured in the experiment is in disagreement with
previous numerical estimates. Another open issue is related to the fact that
only a small fraction of electrons take part in the localized state. It is
clear that an accurate microscopic description of the experimental situation
is necessary to account for all these features.  In our study we include the
most important details such as an accurate screening and the effect of the
finite size of the wire to correctly describe and understand the physics
underlying the experiment.

Throughout the paper we use units of the effective Bohr radius
$a^\star_0 = \frac{\hbar^2 \epsilon}{m^\star e^2}$ for length and the
effective Rydberg $Ryd^\star = \frac{e^2}{2\epsilon a_0^\star}$ for energy
where $\epsilon$ is the dielectric constant of the embedding medium and
$m^\star$ is the effective electron mass. 

The paper is organized as follows: in
Sec.~\ref{unscreened_coulomb_interactions} we present results for a quasi
one dimensional electron gas (1DEG) with long range ($1/x$) interactions for
different densities $\rho=\frac{1}{2r_s}$ and thicknesses.  We carefully
study the liquid-to-quasi-crystal crossover by varying the Wigner-Seitz
radius $r_s$ which sets the relative importance of the kinetic energy and
the interaction.  By means of QMC techniques, we also study the charge
compressibility and the spin susceptibility in order to analyze the
interplay between the charge and spin properties of the wire.  The effect of
the wire's thickness on the crossover and the spin properties is taken into
account by performing simulations with three different wire widths.  In
Sec.~\ref{screened_interactions} we compare the unscreened $1/x$ potential
with an interaction screened by a metallic plane.   In
Sec.~\ref{experiments} we interpret the localization transition found in the
series of two wire tunneling experiments\cite{auslaender-science0,
soft-confinement-prl,soft-confinement-prb,auslaender-solid-state-comm,
auslaender-science1,yacoby,steinberg-tunneling-depleted-top-wire} by
studying the evolution of liquid-to-crystal correlations in a finite wire
with interactions effectively screened by another parallel wire.  
We make a comparison between the finite system and the corresponding homogeneous
infinite system interacting with the same potential.  We also show the agreement
between our model and the experiment.  Finally  in Sec.~\ref{conclusions} 
we summarize our results and comment on possible refinements 
to our calculations.

\section{Unscreened Coulomb interactions}
\label{unscreened_coulomb_interactions}

We study a system of electrons interacting via the Coulomb ($1/x$) potential
which are confined to one dimension by a harmonic potential in the
transverse direction $V(r_\perp) = \frac{r_\perp^2}{4b^4}$, where b tunes
the thickness of the wire.  This system was previously studied using QMC by
Casula \emph{et al.} and here we follow the conventions used in that
work.\cite{casula-1d}  We integrate over the transverse degrees of freedom, 
which is a good approximation when the density of
electrons in the wire is low ($r_s \gg \pi b / 4$), and hence the
longitudinal energy scale is small compared to the excitation energies
related to the perpendicular motion.  This integration yields an effective 
one dimensional interaction: $ V_b(x)
= \frac{\sqrt{\pi}}{b} \exp\left(\frac{x^2}{4b^2}\right)
\textrm{erfc}\left(\frac{|x|}{2b}\right),$ which has a long range $1/x$
tail.  The thickness $b$ of the wire controls the short-range behavior of
the potential, which is finite at the origin ($V(0) = \sqrt{\pi}/b$).  
Since the crossover between the short
and long range behavior is at $x \approx b$, for smaller $b$ the
repulsion is stronger as the particles approach each other.

In this work we have chosen to study three different thicknesses, $b=1$, $0.1$,
and $0.0001$.  The first two values correspond to typical experimental
thicknesses for semiconductor quantum wires, whereas the last one is chosen
to explore the ultrathin limit as studied analytically by
Fogler\cite{fogler-coulomb-tonks-prl,fogler-coulomb-tonks-prb} and
experimentally realized in carbon nanotubes placed on $\textrm{SrTiO}_3$
substrates.\cite{ultrathin-nanotube,ultrathin-nanotube2}  For each value of
$b$, the density in the wire $\left(\frac{1}{2r_s}\right)$ is varied,
allowing for the interaction strength to change and the corresponding ground
state properties are computed.

We use diffusion Monte Carlo (DMC) and lattice regularized diffusion Monte
Carlo (LRDMC)\cite{casula_lrdmc} methods to project 
the initial variational ansatz to the
ground state $|\Psi_0\rangle$.  These methods are particularly suited to the
simulation of one dimensional fermions.  Indeed, the well known ``sign
problem'' does not affect these calculations as the nodes are fully
determined by the points of coincidence between the electrons and therefore
are exactly included in the trial wave function $|\Psi_T\rangle$. Since the final
DMC or LRDMC distribution is the product of the true ground state and the
trial wave function, some observables such as the density and the structure
factor are determined using the forward walking
technique\cite{forward-walking, sorellaFW} in order to generate unbiased
expectation values.

We simulate an unpolarized wire with $N$ electrons subject to periodic
boundary conditions (PBC).  The trial wave function is written in the
Slater-Jastrow form
\begin{equation}
\Psi_T =  D^\uparrow D^\downarrow  \exp \left(-\sum_{i<j} u(x_{ij}) \right),
\label{homo_wave_function}
\end{equation} 
where the Slater determinants for up and down spin electrons read
\begin{equation}
D^\sigma(x^\sigma_1, \ldots, x^\sigma_{N^\sigma}) = 
   \prod_{\substack{1 \le i < j \le N^\sigma}}
   \sin \left( \frac{G}{2} (x^\sigma_i-x^\sigma_j) \right),
\end{equation}
with $G=2 \pi / L$, and $L=2 r_s N$ the length of the simulation cell.  We
follow Ref.~\onlinecite{gaskell} to determine the Jastrow function $u(x)$.
Its Fourier components are
\begin{equation}
2 \rho \tilde{u}(k)= -S_0(k)^{-1}+\sqrt{S_0(k)^{-2}+ 2 \rho \tilde{V}_b(k)/k^2}, 
\label{u_rpa}
\end{equation}
with $S_0(k)=(k/2k_F)\theta(2k_F-k)+\theta(k-2k_F)$ the structure factor of
a noninteracting 1DEG, $\rho = \frac{1}{2r_s}$ the density, and
$\tilde{V}_b(k)$ the Fourier transform of $V_b(x)$.  To reduce the finite
size effects in our simulation we use the Ewald technique to sum our
potential as discussed in detail in Ref.~\onlinecite{casula-1d}. This
approach has been used to study the infinite wire with the long range
potential $V_b(x)$, and also the screened potentials described in the next
Sections.  In the latter case, the sum over the images has been done
numerically in the real space as the potentials have a shorter range.
 
\begin{figure}[!ht]
\centering
\includegraphics[angle=-90,width=\columnwidth]{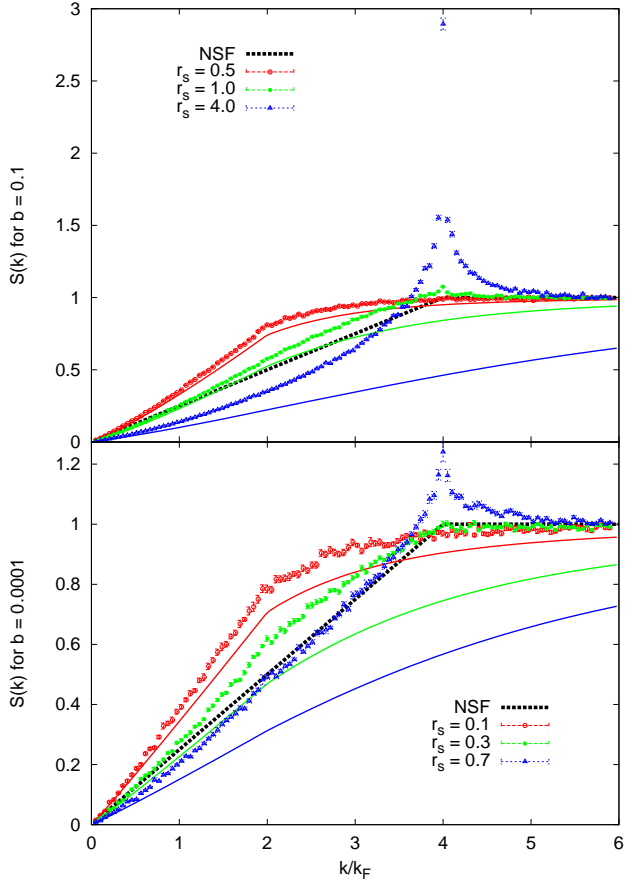}
\caption{(Color online) Structure factor for $b=0.1$ (upper panel), and
$b=0.0001$ (lower panel), computed for a system with $78$ electrons. 
The QMC (points) and MSA (solid lines) structure factors are reported
for different densities ($r_s$). Also the noninteracting spinless fermion (NSF)
structure factor is drawn (solid black line) for comparison.} 
\label{sk}
\end{figure}

\begin{figure}[!ht]
\centering
\includegraphics[angle=-90,width=\columnwidth]{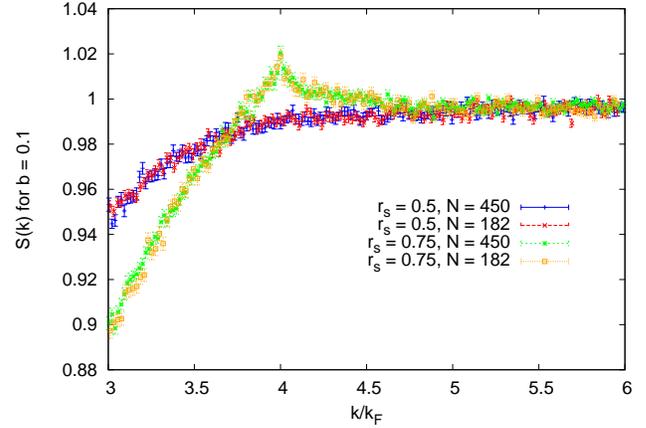}
\caption{(Color online) Detail for the structure factor near $4k_F$ for $b=0.1$, 
computed for $N=182$ and $N=450$ at two densities ($r_s=0.5$ and $r_s=0.75$) in the 
proximity of the crossover from a liquid to a quasi-crystal.}
\label{sk4kf}
\end{figure}

To reveal the presence of charge ordering in the system, we first analyze
the static structure factor $S(k)= \frac{1}{N}\left<\rho(-k)\rho(k)\right>$,
where $\rho(k) = \sum_j e^{ikr_j}$ are the Fourier components of the
electron density.  At high density the structure factor is very similar to
the mean spherical approximation (MSA)\cite{gold-rpa-msa} 
prediction $S_{\textrm{MSA}}(k) = S_0(k)/(1 + 2 \rho \tilde{u}(k) S_0(k))$ 
as expected (see Fig.~\ref{sk}), since in the limit $r_s \rightarrow 0$ the MSA becomes
exact.\cite{casula-1d}  Specifically, there is no peak at $4 k_F$ up to
$r_s=0.5$ ($r_s=0.2$) for $b=0.1$ ($b=0.0001$), namely there are no
correlations with the mean interparticle spacing (Fig.~\ref{sk4kf}).  As the
density decreases, a peak develops at $4 k_F$.  This peak is a necessary
feature for a one dimensional quasi Wigner crystal and it is absent in the
MSA prediction which has no structure at $4 k_F$.  For $b=0.1$ we carried
out simulations with up to $450$ particles for $r_s=0.5$ and $r_s=0.75$, to
check the convergence of the $S(k)$ in the liquid regime close to the onset
of the $4 k_F$ charge correlations (Fig.~\ref{sk4kf}). 

\begin{figure}[!ht]
\centering
\includegraphics[angle=-90,width=\columnwidth]{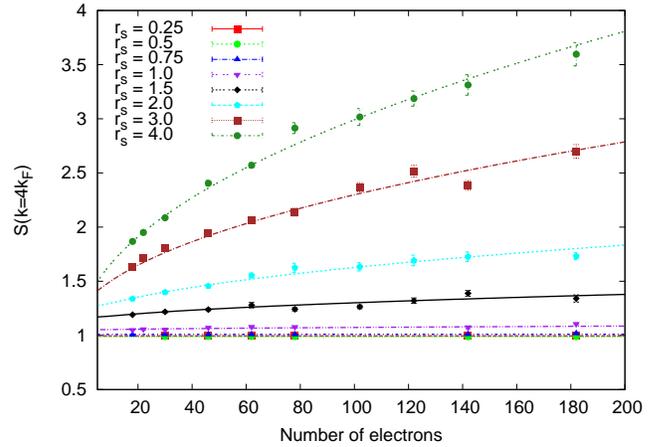}
\caption{(Color online) Scaling of the $4 k_F$ component of the structure
factor with respect to the number of particles. The scaling is reported for
various densities with $b=0.1$. The lines are the best fit of the function 
in Eq.~\ref{schulz_structure_factor} given by the LL theory.} 
\label{SkScaling}
\end{figure}

The scaling of the height of the $4 k_F$ peak of $S(k)$ with the number of
particles (reported in Fig.~\ref{SkScaling} for $b=0.1$) highlights the
features of a liquid-to-quasi-crystal crossover.  When the peak is absent
there is no significant dependence of the $S(4k_F)$ value as a function of
system size, however where there is a peak in the structure factor at $4
k_F$, its scaling is sub-linear, signaling a quasi-long range order (linear
scaling would indicate a true Wigner crystal).  The points in
Fig.~\ref{SkScaling} are fit very well by a functional form obtained from
the charge-charge correlation function\cite{casula-1d-correlations} derived
by Schulz\cite{schulz} in the LL framework with long range interactions, 
\begin{equation}
\int_{c_0}^{L} \textrm{d}x ~ \exp(-i4k_F x)\left\langle \rho(0) \rho(x) \right\rangle  = 
a L \exp(-4c\sqrt{\log L}) + b,
\label{schulz_structure_factor}
\end{equation}
where we explicitly include the dependence on the system size $L$ by taking the Fourier 
transform over the simulation cell. The short-distance cutoff $c_0$ is introduced 
because the LL theory provides only the asymptotic behavior for $\left\langle \rho(0) \rho(x) \right\rangle$.
Further logarithmic corrections could be included\cite{giamarchi} in Eq.~\ref{schulz_structure_factor},
but we take just the leading order expansion, which should be the most relevant for the system sizes
computed here.  One would need much larger systems which are beyond our current numerical 
capabilities to resolve further corrections.
The bosonization formalism gives a parameter dependent scaling for the $4 k_F$ component
of $\left\langle \rho(0) \rho(x) \right\rangle$ 
which is left undetermined in the LL theory, and depends on the details of the interaction.  
At high densities there is no peak in the structure factor 
and the electron gas is liquid (Fig.~\ref{sk4kf}). Consequently, there is 
no finite size dependence at $4k_F$ and the parameter $a$ 
undetermined in the LL theory is zero.

We also determine the charge compressibility $\chi_\rho$ and 
the spin susceptibility $\chi_\sigma$ of the electron gas 
by using two techniques in our calculations. The first is to apply 
the definition of those quantities as the reciprocal of the
second derivative of the total energy with respect 
to the density $r_s$ or the polarization $\zeta$.  
Our QMC calculations provide measurements of the total energy, so these derivatives can be
taken by fitting our data with a suitable functional form. 
The error in such a determination comes from both the statistical 
uncertainty in the calculations and the constraint represented 
by the choice of the fitting function. As a technical detail, it is also necessary to 
extrapolate the energy to the thermodynamic limit which can be a costly proposition.
Moreover a functional form which includes the dependence on both density 
and polarization has not been provided yet.\cite{shulenburger} 
Nevertheless, we use the parametrization in Ref.~\onlinecite{casula-1d}, 
which holds for a system with fixed polarization $\zeta$ 
and depends only on $r_s$ to compute the charge compressibility
and validate the second method to evaluate $\chi_\rho$ and $\chi_\sigma$.

The other method we use to compute these quantities is to calculate the momentum resolved
excitation energies of the system, and exploit the sum rules which \emph{exactly} 
relate the collective modes of the long wavelength spectrum 
with $\chi_\rho$ and $\chi_\sigma$. Gold and Calmels\cite{gold-excit} found that
\begin{align}
  \label{charge-excitation-spectrum}
  \omega_\rho(k\rightarrow 0) &= v_F|k|\sqrt{\rho_F V(k\rightarrow 0) + \frac{\chi_0}{\chi_\rho}}, \\
  \label{spin-excitation-spectrum}
  \omega_\sigma(k\rightarrow 0) &= v_F|k|\sqrt{\frac{\chi_0}{\chi_\sigma}},
\end{align}
where $\omega_\rho(k)$ ($\omega_\sigma(k)$) is the energy of 
the lowest charge (spin) excitation with momentum $k$, 
$\rho_F$ is the density of states of the free electron gas at the Fermi energy, 
and $\chi_0 = 16r_s^3/\pi^2$ is its compressibility.

In order to find out the lowest energy states of a given momentum $k$ we employ a method proposed 
by Ceperley and Bernu,\cite{ceperley-cfmc} which is a generalization of the transient estimate 
used in the projection Monte Carlo (DMC or LRDMC) framework. This method is based on the idea 
that it is possible to compute the excitation spectrum of a system in a direct and variational way 
by projecting the initial basis functions to their lowest energy state with the given symmetry.
In our case the basis set is the Feynman
ansatz,\cite{feynman-helium-excitations} i.e. $\rho(k) | \Psi_0 \rangle$ $\forall k$ for the charge
excitations and $\sigma(k) | \Psi_0 \rangle$ $\forall k$ for the spin excitations, where 
$ \sigma(k) = \sum_j \sum_\sigma  \sigma e^{ikr^\sigma_j}$ is the Fourier transform of the 
spin density. In the following we assume to work with the charge excitations, but the 
same applies for $\sigma(k)$.
Since the basis set is orthogonal, the method in Ref.~\onlinecite{ceperley-cfmc} is greatly simplified, 
as every $k$ component is decoupled. For each $k$, we have to calculate
\begin{equation}
\frac{ \left<\Psi_0\right| \hat{\rho}(k,\tau) \hat{H} \hat{\rho}(-k,0) \left|\Psi_0\right> }
     { \left<\Psi_0\right| \hat{\rho}(k,\tau) \hat{\rho}(-k,0) \left|\Psi_0 \right>} = 
  \frac{\sum_i \epsilon_k^i A_k^i e^{-\tau (\epsilon_k^i-E_0)}}
       {\sum_i A_k^i e^{-\tau (\epsilon_k^i-E_0)}},
\label{diagonal}
\end{equation}
where $\hat{\rho}(k,\tau)$ is written in the Heisenberg representation with imaginary time evolution, 
$\left|\Psi_k^i\right\rangle$ is the $i$'th excited state with momentum $k$, $\epsilon_k^i$ is its energy,
$A_k^i=\left| \langle \Psi_k^i | \rho(-k) | \Psi_0 \rangle \right|^2$ 
is the spectral weight of the eigenvalue expansion,
and $E_0$ is the ground state energy.
For large $\tau$ the ratio in the above Equation 
will converge to the lowest energy $\epsilon_k^0$ of a given $k$, 
provided $A_k^0$ is non zero. Another limitation is given 
by the exponentially small denominator, which will 
exponentially increase the statistical noise of the estimate 
as the projection time increases. Both the numerator 
and denominator in Eq.~\ref{diagonal} are 
evaluated by means of the forward walking\cite{forward-walking,sorellaFW} procedure based on the
DMC or LRDMC sampling. Indeed, for large enough $\tau$ the LHS of Eq.~\ref{diagonal} can be rewritten as
\begin{equation}
\frac{ \int \textrm{d} r_1 \textrm{d} r_2 ~ \rho(-k) G(r_1,r_2,\tau) E_L(k,r_2) \rho(k)  P(r_2)}
     { \int \textrm{d} r_1 \textrm{d} r_2 ~ \rho(-k) G(r_1,r_2,\tau) \rho(k)  P(r_2)  },
\label{diagonal_qmc}
\end{equation}
where $E_L(k,r)= \frac{H \rho(k) \Psi_T(r)}{\rho(k) \Psi_T(r)}$ 
is the local energy of $\rho(k) | \Psi_T \rangle$,
$P(r)=\Psi_T(r) \Psi_0(r)$ is the QMC mixed distribution, and 
$G(r_1,r_2,\tau)=\Psi_T(r_1) \langle r_1 | e^{-\tau H} | r_2 \rangle / \Psi_T(r_2)$ is the 
importance sampled Green's function. 

\begin{figure}[!ht]
\centering
\includegraphics[angle=-90,width=\columnwidth]{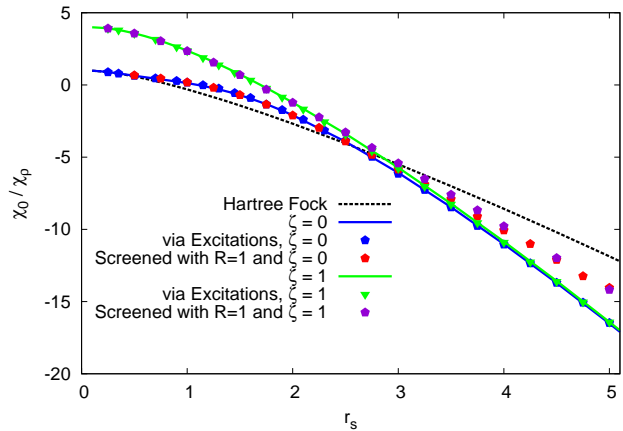}
\caption{(Color online) Inverse charge compressibility $\chi_0/\chi_\rho$ of the
unpolarized and fully polarized wire for $b=0.1$, 
with both screened ($R=1$) and unscreened interactions.
Also the HF (dashed back line) charge
compressibility is reported for the unpolarized wire.
The solid lines are obtained from the second derivative of the energy
parametrization, while the points are evaluated through the
charge excitations as explained in Sec.~\ref{unscreened_coulomb_interactions}.
}
\label{charge_comp}
\end{figure}

Because the excitation energies $\omega(k)=\epsilon_k^0-E_0$ 
are computed relative to the ground state energy $E_0$, 
there is a cancellation of errors since the sample generated to compute $E_0$ and $\epsilon_k^0$ is the same.
Therefore a modest size calculation is enough to get converged energies.  
The convergence with the propagation time can be more difficult to obtain.
However, for the long wavelengths $\rho(k)\left|\Psi_0\right>$ is a good approximation 
to the lowest excited state with momentum $k$ and the energies 
can be determined easily with a short projection time $\tau$.  
When the small $k$ range of energies is fit to the form in
Eqs.~\ref{charge-excitation-spectrum} and \ref{spin-excitation-spectrum},
$\chi_\rho$ and $\chi_\sigma$ are determined.
The results for the charge compressibility obtained with this method agree 
with the second derivatives of the total energy in all cases we have made the comparison,
as is shown in Fig.~\ref{charge_comp}.

\begin{figure}[!ht]
\centering
\includegraphics[angle=-90,width=\columnwidth]{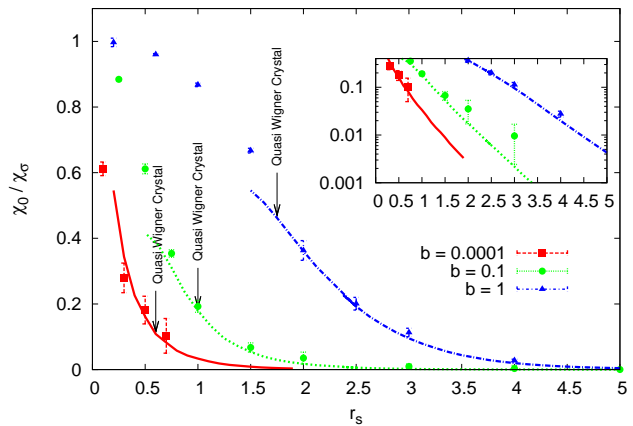}
\caption{(Color online) Inverse spin susceptibility $\chi_0/\chi_\sigma$ for different
thicknesses. The dependence on $r_s$ is shown. The points are the QMC
calculations, while the lines are the WKB estimates. The arrows indicate the
liquid-to-quasi-Wigner-crystal crossover.} 
\label{spin_susc}
\end{figure}

The knowledge of $\chi_\rho$ and $\chi_\sigma$ can shed more light on the properties of the
liquid-to-quasi-crystal crossover. By looking at
the charge compressibility (Fig.~\ref{charge_comp}), it is apparent that the
role of the electron correlation is becoming increasingly important in the proximity of the
crossover, where there is significant discrepancy between the Hartree-Fock (HF) 
and QMC values of $\chi_\rho$.
In particular, the correlation makes the system
\emph{softer} than the HF, which is consistent with a more pronounced localization of the
electrons. At even lower densities the charge compressibility of the
unpolarized system is approaching that of a fully polarized (or spinless
fermion) gas. The difference between the two is going exponentially to zero,
and they almost overlap for $r_s > 4$ (with $b=0.1$). This means that the energy of the
spin excitations is getting smaller and smaller as the density decreases. This
feature is revealed by the inverse spin susceptibility $\chi_\sigma$.
The $\chi_0/\chi_\sigma$ ratio is plotted in Fig.~\ref{spin_susc}.
This value becomes exponentially small at low
densities, where it is difficult to get a statistically accurate QMC estimate,
since the sampling of the spin is ``frozen'' by the presence of quasi nodes 
(pseudo nodes) between unlike spin electrons.\cite{casula-1d} The strong interaction 
makes the electrons to repel each other at short-range, and the corresponding wave function
is very small at the coalescence points of electrons with opposite spin.
Consequently the spin flip rate in the QMC sampling becomes small, and the efficiency 
decreases. However the charge properties do not seem to be affected by this slowing-down. 
The physical reason for the quasi nodes will become even more apparent in Sec.~\ref{screened_interactions},
when we will discuss the Tonks-Girardeau physics of the screened wire.

In the low density regime where exact Monte Carlo sampling becomes difficult
the WKB approximation is useful for determining the dynamical properties of the
electron gas. Following the example of Matveev\cite{matveev-conductance-of-quantum-wire} we use the
WKB approximation to determine the rate at which two electrons exchange by
calculating the energy barrier that they must overcome.
Although fluctuations prevent the formation of a Wigner crystal, the 
equilibrium positions of the electrons are assumed to be equally spaced with periodicity $2r_s$. 
Central to the accuracy of this approximation is the fact that at low
densities the tunneling is dominated by the effect of the potential, and
the statistics can be ignored.
Furthermore, all electrons are treated as uncorrelated except for a single pair which is allowed to exchange.
In contrast to Matveev's approach we assume that the other electrons are distributed 
about their equilibrium positions according to the harmonic approximation with a 
Gaussian spread instead of being fixed delta function point particles. 
Taking the initial positions of the two exchanging electrons to be at $x = 0$ and $x = 2r_s$, 
they feel a static potential given by
\begin{equation}
\label{wkb-potential}
V_{WKB}(x) = \sum_{n\neq0,1} \int_{-\infty}^{\infty} \rho(y) V(x-2nr_s+y) dy, 
\end{equation}
where $\rho(y) = \sqrt{\alpha/\pi}\exp(-\alpha y^2)$ is the equilibrium charge
density of the non exchanging electrons and $V(x)$ is the interparticle
potential. The harmonic approximation 
gives $\alpha = \sqrt{m \frac{\partial^2 W(x)}{\partial x^2}}$,
where $W(x)$ is the potential at a given lattice site due to an infinite array of electrons
spaced as $2r_s$.

At low densities the electrons behave as a spin chain obeying the
Heisenberg Hamiltonian were the spin flips are mediated by an exchange of nearest neighbor
electrons, so the spin susceptibility can be determined from the energy barrier computed
within the WKB approximation by analogy with the Heisenberg Hamiltonian in 1D as shown by 
Matveev.\cite{matveev-conductance-of-quantum-wire}  The spin velocity of the equivalent
Heisenberg spin chain can be found from the Bethe ansatz solution\cite{bethe-ansatz1, bethe-ansatz2}, 
yielding $v_\sigma = \pi J r_s / 2$ where $J$ is the size of the energy barrier in the WKB approximation.  
This gives the susceptibility through Eq.~\ref{spin-excitation-spectrum}.

Where the density is large enough that QMC reliably samples 
the spin exchanges the spin susceptibility computed using the forward walking techniques 
agrees well with the WKB estimate only after the smearing of the electron sites 
given by the harmonic approximation. It is therefore 
important to use the potential in Eq.~\ref{wkb-potential} to have an accurate estimate 
of the exchange at intermediate densities. This agreement and the fact 
that the dynamical many-body corrections to the WKB estimate are very small
at low density\cite{klironomos}
justify the use of WKB for dilute systems where it is difficult to extract 
information from the QMC calculations.  In addition, the exponential decay 
of $v_\sigma$ versus $\sqrt{r_s}$ obtained in this way is in agreement with previous results 
\cite{matveev-conductance-of-quantum-wire, fogler-exchange-quantum-rings, fogler-spin-exchange}
for potentials where they can be compared.

Fig.~\ref{spin_susc} summarizes our findings for the unscreened wire.
The liquid-to-quasi-crystal crossover is shifted to higher densities for
thinner wires, while the spin susceptibility is always significantly different
from zero in the crossover region for the values of the confinement taken into
account. The smallest $b$ we studied ($b=0.0001$) corresponds to one of the
thinnest confinements realized experimentally.\cite{ultrathin-nanotube,ultrathin-nanotube2} The
spin exchange is still sizable in the crossover region due to the not-so-long
localization length of the electrons and not-so-thin width of the wire.
Therefore, in our study we did not find any signature of the Coulomb-Tonks gas
phase in between the liquid and quasi Wigner crystal, which was claimed by
Fogler for ultrathin wires.\cite{fogler-coulomb-tonks-prb}  
However, the structure factor plotted in Fig.~\ref{sk} reveals the tendency for electrons
to approach the noninteracting spinless fermion behavior (the limit where the
Coulomb-Tonks gas picture holds) as the wire width decreases. The fundamental difference
with respect to the noninteracting spinless picture is the pronounced peak at $4 k_F$,
which characterizes the Coulomb long-range interactions at low density.

\section{Screened interactions in gated wires}
\label{screened_interactions}

The primary interest of this paper is to model 
a quantum wire formed in a semiconducting nanodevices. In that case there is almost
always a metallic gate that screens the long range ($1/x$) potential. To see
the changes that such a gate would cause, we introduce 
a perfectly conducting metal plane parallel to
the wire located a distance $R$ away. Using the electrostatic method of images
the potential is constructed by assuming that a wire is placed at a distance 
$2R$ from the original one with the same particle distribution but opposite sign.
The equation for this potential is
\begin{align}
\label{gate-screened}
V(x) &= \int \int d\,\vec{r} d\,\vec{r}^{\,\prime}  \frac{\rho_b(\vec{r})\rho_b(\vec{r}^{\,\prime})}
        {\sqrt{(\vec{r}-\vec{r}^{\,\prime})^2 + x^2}} - \nonumber \\ \nonumber
     & \int \int d\,\vec{r} d\,\vec{r}^{\,\prime} \frac{\rho_b(\vec{r})\rho_b(\vec{r}^{\,\prime})}
        {\sqrt{(\vec{r}-\vec{r}^{\,\prime}-2\vec{R})^2 + x^2}} \\ 
     &= V_b(x) - V_{int}(x,R)
\end{align}
where $\vec{r}$ and $\vec{r}^{\,\prime}$ are transverse vectors, 
$\rho_b(\vec{r}) = \frac{1}{b\sqrt{2\pi}} \exp\left(-\frac{r^2}{2b^2}\right)$ is the ground state charge 
distribution of a two dimensional harmonic oscillator with the wire's confining potential: 
$V_{\textrm{wire}}(r) = \frac{r^2}{4b^4}$.  The first integral gives the effective unscreened
inter-particle potential $V_b(x)$ described in the previous section
and the second one is the potential due to the image charge on the screening wire: $V_{int}(x,R)$.

\begin{figure}[!ht]
\centering
\includegraphics[angle=-90,width=\columnwidth]{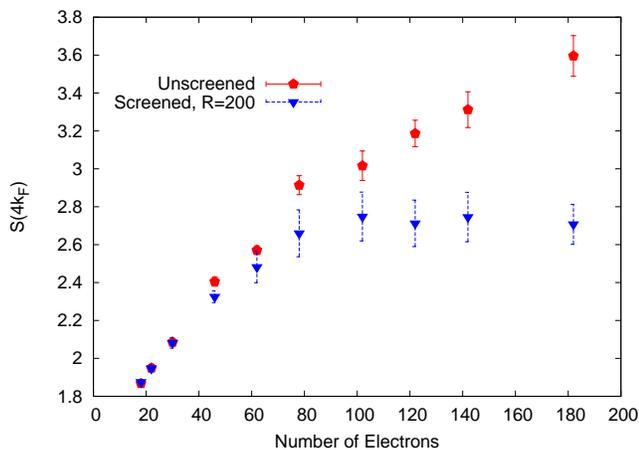}
\caption{(Color online) Scaling of the $4 k_F$ component of the structure
factor with respect to the number of particles for $b=0.1$, and $r_s=4$. 
For comparison, the scaling is reported for the unscreened $V_b$ interaction, and 
the screened potential in Eq.~\ref{gate-screened} with $R = 200$.} 
\label{SkScaling-scr}
\end{figure}

The quasi Wigner crystal correlations derived by Schulz\cite{schulz} apply only 
when the interaction is long range ($1/x$).  In the case of
the screened interaction above the potential decays as $4R^2/x^3$ at large
distances, so a simple scaling argument shows that the Wigner crystal correlations 
should be absent at very low densities.  Indeed, if $r_s > 8R^2/\pi$ the typical kinetic energy of 
the electrons, the Fermi energy $E_F$, 
is larger than the potential energy computed at the mean interparticle distance ($2 r_s$).
At these low densities Matveev\cite{matveev-conductance-of-quantum-wire}
has pointed out that it is possible to map the screened short-range 
interaction into a repulsive contact potential
\begin{equation}
\label{delta-fcn-interaction}
 V(x) = U\delta(x),
\end{equation}  
where the constant $U$ is chosen so the delta function potential and the screened
one have equal transmission coefficients.  On the other hand, in the density range 
$1 \ll r_s <  8R^2/\pi$ the $1/x$ shoulder of the potential can induce $4 k_F$ correlations,
which are strong but not strong enough to stabilize any sort of quasi-order. Calculations of the
finite size scaling of the $4k_F$ peak of the structure factor for $b=0.1$ and $R=200$ show the 
saturation of its height for $N \gtrsim 100 $, and so demonstrate the absence
of the quasi Wigner crystal correlations when screening is introduced despite quite a large distance
to the metallic gate (Fig.\ref{SkScaling-scr}). Only in the limit of $R \rightarrow \infty$ does one recover
the unscreened potential and the possibility for a quasi long-range charge order. 

The lack of the quasi Wigner crystal state does not change the crossover to 
the spinless fermion physics present in the unscreened system. 
Even though the quasi long-range charge order is absent at low densities, 
the charge compressibility still approaches that of a gas of spinless fermions as the density 
decreases.   This approach can seen in Fig.~\ref{charge_comp}.  It is therefore clear that the spin
crossover does not depend on the long-range correlations.  In fact, this crossover 
can be reproduced by a system of electrons interacting via the delta function 
interaction in Eq.~\ref{delta-fcn-interaction} where the constant U is large, an
interaction that has no long-range piece whatsoever.

The low density limit with screened interactions is particularly interesting as the
screening introduces a new feature.  At low densities
the electron-electron repulsion at short range makes exchanges between electrons
virtually impossible, corresponding to the limit $U\rightarrow \infty$.
As a result for the ultrathin wire with strong screening ($b \ll 1$ and $r_s \gg 8R^2/\pi$), 
the mapping of the interaction to the potential in Eq. 12 becomes \emph{exact}.  
In this situation not only do the electrons behave as spinless fermions, but the charge velocity
approaches that of \emph{noninteracting} spinless fermions ($v_\rho= 2 v_F$).  This
is analogous to the case of bosons with infinite repulsive contact interactions, (or 
impenetrable particles) where the system can be mapped into a noninteracting Fermi
gas.\cite{girardeau}  The impenetrable Bose system is often called a Tonks-Girardeau
gas.  In our case the situation is analogous, namely the fermions become impenetrable due
to an effective infinite contact repulsion, and so they behave as they were noninteracting and spinless.
We refer to this behavior as Tonks-Girardeau regime. One of its features is the 
presence of nodes in the wave function at the coalescence of unlike spin pairs. 
This is the extreme case when the pseudo nodes that complicate the ergodicity 
of Monte Carlo calculations at low density as reported in
Sec.~\ref{unscreened_coulomb_interactions} become actual nodes.

\begin{figure}[!ht]
\centering
\includegraphics[angle=-90,width=\columnwidth]{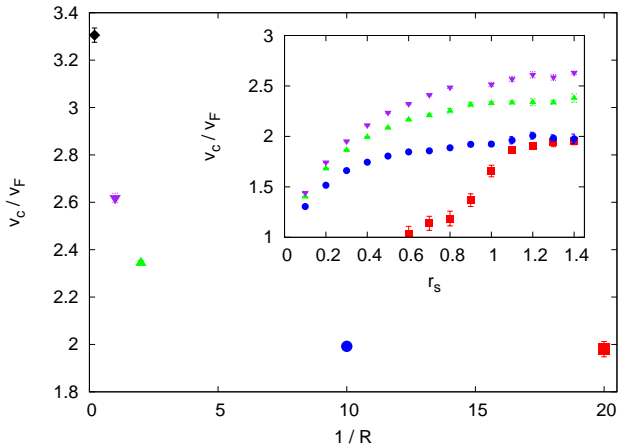}
\caption{(Color online) Asymptotic large $r_s$ values of the charge velocity in units of $v_F$ vs. inverse 
screening length for ultrathin wire ($b=0.0001$) from $R=0.05$ to $R=5$. In the inset we report the 
full dependence of the charge velocities on $r_s$ at different $R$.} 
\label{tonks-girardeau-graph}
\end{figure}

While this effect has been discussed in the literature,
\cite{matveev-conductance-of-quantum-wire,fogler-coulomb-tonks-prb,fogler-coulomb-tonks-prl}
our work provides quantitative predictions for the onset of the noninteracting spinless behavior.  
Fig.~\ref{tonks-girardeau-graph} shows the charge velocity in the limit of low density 
for different values of the screening in the thinnest wire we studied ($b=0.0001$).  
We found that in order for the Tonks-Girardeau behavior to manifest itself, 
the distance to the gate $R$ must be less than $0.1$ and the density must be lower than $r_s = 1$.
For $R$ larger than $0.1$, at low density the charge velocity does not converge 
to the noninteracting spinless fermion limit ($2 v_F$), but saturates at a larger value.

\begin{figure}[!ht]
\centering
\includegraphics[angle=-90,width=\columnwidth]{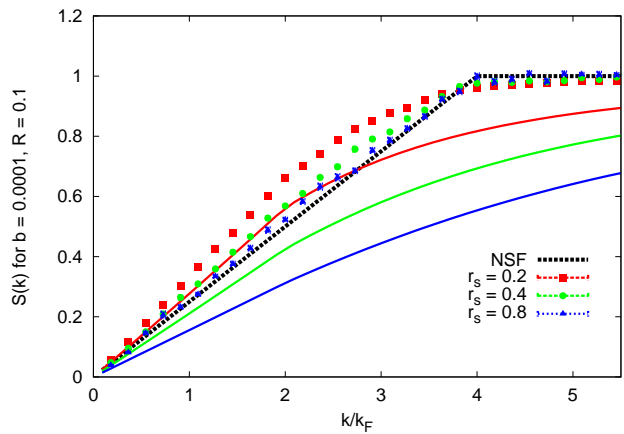}
\caption{(Color online) Static structure factor for the screened wire with $b=0.0001$ and $R=0.1$, 
plotted for three values of the density, $r_s=0.2$,$0.4$ and $0.8$.  The solid lines
correspond to the MSA prediction for each density, and the black line is the structure factor for
noninteracting spinless fermions (NSF). } 
\label{sk-scr}
\end{figure}

It is possible to see the transition of the screened electron gas
to the noninteracting spinless fermion behavior more directly by analyzing the static structure factor,
as was done in the unscreened case. In Fig.~\ref{sk-scr}, the $S(k)$ is plotted at different densities
for the ultrathin wire with $b=0.0001$ and gate located at $R=0.1$ from the wire. Contrary to the 
case of the unscreened wire (Fig.~\ref{sk} lower panel), at low densities the peak at $4k_F$ is absent
and the structure factor approaches that for noninteracting spinless fermions quite closely. Notice that
at the same time the charge velocity approaches the value of $2 v_F$ (see Fig.~\ref{tonks-girardeau-graph}).

The same study was repeated for the wire with $b=0.1$.  Here the short-range behavior of the
potential is much less repulsive than in the $b=0.0001$ case and the same
value $R$ for the screening. The result of this is that the charge velocity does not converge to $2 v_F$
even for a gate as close as $R=0.1$, which equals the width of the wire and thus represents
the geometric limit of validity for the uncorrelated inter-wire interaction.
Therefore for $b=0.1$ and thicker wires, whose widths are realizable in semiconducting
nanostructures, we did not find the Tonks-Girardeau behavior in our calculations.

\section{Localization transition in two parallel G\lowercase{a}A\lowercase{s} wires}
\label{experiments}

Quasi one dimensional systems can be realized in GaAs/AlGaAs
heterostructures by means of various techniques.  One such technique being
cleaved edge overgrowth, which has been applied recently to build an
experimental setup with two parallel wires so that it is possible to observe
momentum resolved tunneling from one to the
other.\cite{auslaender-science0,soft-confinement-prl,soft-confinement-prb, 
auslaender-solid-state-comm,auslaender-science1,yacoby,
steinberg-tunneling-depleted-top-wire}
In this series of experiments both the energy of the tunneling electrons and
their momentum could be tuned by changing the relative chemical potential
and the applied magnetic field.  This setup allows the dispersion relations
of each wire to be probed in a quite straightforward manner.  Steinberg et
al.\cite{steinberg-tunneling-depleted-top-wire} further explored how this
tunneling is affected by a gate that depletes the density of the electrons
in the upper wire.  They found that as the density is decreased there is a
marked transition in the tunneling interpreted as a transition from a liquid
to a localized state.

In the experiment, the center to center distance between the two wires is R
= 31nm. The upper wire is  $2\,\mu m$ long and 20 nm wide.  It is the probe
to study the electron localization. The electrons tunnel from the lower
wire, which has a width of 30 nm and is taken to be infinitely long. This is
also a screening medium for the upper wire. The system is fabricated out of
GaAs for which $\epsilon = 13.1$ and the effective electron mass is $m^\star
= 0.067\,m_e$. This gives an effective Bohr radius $a_0^\star =
\frac{\epsilon \hbar^2}{m^\star e^2} \approx 10\, nm$.  For the experiment
in question the electron density in the lower wire is around $60 \mu m^{-1}$, 
which corresponds to $r_s = 0.83$ in $a_0^\star$ units, while in
the upper wire the density is varied by tuning the gate voltage $V_G$. The
effect of $V_G$ on the lower wire is very
small\cite{steinberg-tunneling-depleted-top-wire} and can be neglected.

The results presented in the previous section offer an avenue to explore the
role of the electron correlation in the transition observed in the
experiment. As the density in the wire decreases the strength of the potential increases
relative to the kinetic energy.  One effect of this increased relative
strength is that exchanges between the electrons are suppressed, causing the
system to crystallize.  To better quantify the
importance of this effect in the experimental system, in this section we
take into account a more realistic potential, assuming the electrons are
screened by the lower wire instead of an infinite metallic gate. To
construct this interaction we neglect the correlation between the wires and
treat the screening effects coming from the electrons in the lower wire
within the linear response theory. We write the potential in Fourier space 
\begin{equation}
\label{wire_screened_interaction}
V(k,R) = V_b(k) + V_{int}(k,R) \chi(k) V_{int}(k,R),
\end{equation}
where $V_b(k)$ and $V_{int}(k,R)$ defined in Eq.~\ref{gate-screened} 
are the intra- and inter- wire potentials respectively. 
$V_{int}(k,R)$ is evaluated by assuming that the thickness of
the two wires is the same (and equal to the upper wire). This significantly
simplifies the form and the calculation of the inter-wire interaction.
$\chi(k)$ is the static density-density response function of the lower wire,
taken in the random phase approximation (RPA):
\begin{equation}
 \chi_{RPA}(k) = \frac{\chi_0(k)}{1 - V_{b^\prime}(k)\chi_0(k)}
\label{chi_RPA}
\end{equation}
where $\chi_0(k) = \frac{1}{\pi k}\ln\left | \frac{k-2k_F}{k+2k_F} \right | $
is the static response function for a one dimensional noninteracting Fermi
gas, and $b^\prime$ is the width of the lower wire.  The experimental
geometry sets the parameters in our quasi one dimensional interaction
$V(k,R)$.  The confinement potential for the upper wire is chosen so that 
the electrons are constrained to be inside the $10 nm$ thick wire.
Specifically, we require the radial root mean squared displacement is equal 
to the lithographic thickness yielding $b=0.707 (\approx 1/\sqrt{2})$ for
the upper wire.  The choice of confinement also agrees
well with the experimental observation that a second mode becomes populated
at $n = 80\mu m^{-1}$.\cite{auslaender-science0}  Similarly, the lower wire's thickness
is given by $b^{\,\prime}=1.061 (\approx 1.5/\sqrt{2})$.  The distance between the wires is 
$R=3.0$, while the Fermi momentum in the RPA response function for the lower 
wire is set by the density $r_s = 0.83$.

Our screened potential in Eq.~\ref{wire_screened_interaction} is similar to
that used by Fiete \emph{et al.},\cite{fiete} who chose a perfect metal
response function which is valid when the screening wire is at very high
densities.  Here we use the RPA which depends on the experimental density of
the lower wire through the value of the Fermi momentum $k_F$.  We notice
that our screened potential equals that in Ref.~\onlinecite{fiete} at $k=2
k_F$ and in the limit of small $k$, namely the long-range tail is the same,
decaying approximately as $1/x^{\frac{5}{4}}$.

\begin{figure}[!ht]
\centering
\includegraphics[angle=-90,width=\columnwidth]{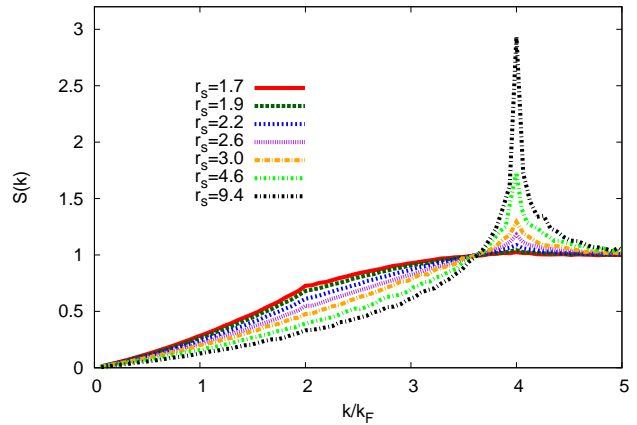}
\caption{(Color online) Static structure factor for a homogeneous wire with
$b=0.707$ interacting with the effective potential in
Eq.~\ref{wire_screened_interaction}, which includes the screening by another
homogeneous wire with $r_s=0.83$, $b=1.061$, and $R=3$.  The structure factor
is plotted for several values of the upper wire density, with $r_s$ ranging 
from $1.7$ to $9.4$.  The calculations have been converged
to the thermodynamic limit, requiring $N=62$ for $r_s \le 3.0$ and $N=78$
subject to periodic boundary conditions for $r_s = 4.6$ and $9.4$.}
\label{sk-wirescr}
\end{figure}

We first analyze the homogeneous system and then explicitly include a
longitudinal confinement in our simulations to quantify the finite-length
impact on the properties of the system, and more closely reproduce the
experimental situation.  
In the homogeneous system of electrons interacting via the potential in
Eq.~\ref{wire_screened_interaction}, 
we observe the appearance of a $4 k_F$ peak in the $S(k)$ around $r_s=2.2$. 
As shown in Fig.~\ref{sk-wirescr}, it is clearly visible for $r_s > 2.6$, whereas
no peak is discernable for $r_s \leqslant 1.9$. This crossover is 
similar to that found for long range $1/x$ interactions. However, the important
difference here is that the quasi long-range order is not present in this case.
Indeed, we have made a systematic study of the scaling with size, 
and the height of the peak converges to a finite value 
in the thermodinamic limit for all densities taken into account.
This behavior is consistent with the decay of the screened interaction 
which is \emph{faster} than $1/x$.\cite{schulz}
Therefore, the crossover is between a high-density liquid
to one with strong $4k_F$ correlations, whose onset 
can be related to the transition occurring in the experimental system.  

The above treatment of the upper wire as infinite and homogeneous can be
improved to resemble the experiments more closely. In the study of the 1DEG
there are strong effects due to any perturbation that breaks the
translational invariance of the system. For instance, Tserkovnyak \emph{et
al.} showed that the asymmetry in the oscillations of the conductance as a
function of the momentum transferred between the two wires can be explained
at the WKB level by having a soft confinement potential for the upper
wire.\cite{soft-confinement-prl}  In a later paper they accurately
determined the functional form of the longitudinal confinement by fitting
its parameters to reproduce the period of those oscillations as a function
of the magnetic field applied to the sample.\cite{soft-confinement-prb}  The
potential that provided a good fit to their data reads
\begin{equation}
V(x) = E_F \left( \frac{2x}{L} \right)^8,
\label{confinement}
\end{equation}
where $E_F$ is the Fermi energy of the upper wire, and $L$ is approximately
1.5 times the lithographic length of the upper wire, namely $L = 300$ in
$a_0^\star$ units.

We used the above potential together with the interparticle potential in
Eq.~\ref{wire_screened_interaction} to study the effect of the confinement
on the transition. Although in principle diffusion Monte Carlo yields an
unbiased ground state energy in one dimension even for a confined system,
(the nodes being exactly determined by the coalescence conditions just as in
the infinite homogeneous wire) in practice it is necessary to improve the
guidance wave function to reduce the variance of our estimates. The Jastrow
factor used in the homogeneous system (Eq.~\ref{u_rpa}) is replaced by a
more sophisticated factor including one-, two-, and three-body terms, fully
optimized by means of the stochastic reconfiguration (SR)
algorithm,\cite{sorellaSR,casulaSR} while the Slater part is kept the same
as in Eq.~\ref{homo_wave_function}.  The one-body Jastrow $\exp(J_1)$ is
needed to localize the electrons in the finite system. It reads
\begin{equation}
J_1 = \sum_{i=1}^N \left(-\alpha x_i^4 -\beta x_i^5\right),
\label{one_body}
\end{equation}
where $\alpha$ is a free parameter and $\beta=\sqrt{E_u} (2/L)^4 / 5$ is
fixed to cancel the contribution of the potential to the local energy at the
leading order in the large distance expansion.  The two-body $\exp(J_2)$ and
three-body $\exp(J_3)$ Jastrow factors are given by
\begin{equation}
J_2=\sum_{(i\sigma)<(j \sigma^\prime)} u_2^{\sigma \sigma^\prime}(x_{ij}),
\label{two_body}
\end{equation}
and
\begin{equation}
J_3= \sum_{(i\sigma),(j \sigma^\prime),(k \sigma^{''})} 
     u_3^{\sigma \sigma^\prime}(x_{ij})~ u_3^{\sigma^\prime \sigma^{''}}(x_{jk}),
\label{three_body}
\end{equation}
where $x_{ij}$ is the interparticle distance. Since the finite system with
screened interactions is dominated by short-range correlations, we chose
$u_n(x)$ to have a simple Gaussian form
\begin{equation} 
u_n^{\sigma \sigma^\prime}(x)=\delta_n^{\sigma \sigma^\prime} 
           \exp \left( - x^2 / \gamma_n^{\sigma \sigma^\prime} \right),
\label{ujastrow}
\end{equation}
with $\delta_n^{\sigma \sigma^\prime}$ and $\gamma_n^{\sigma \sigma^\prime}$
variational parameters.  Energy minimization improves the quality of the
variational wave function and stabilizes the forward walking
estimate\cite{sorellaFW} of the expectation values on the DMC projected state.

\begin{figure}[!ht]
\centering
\includegraphics[angle=-90,width=\columnwidth]{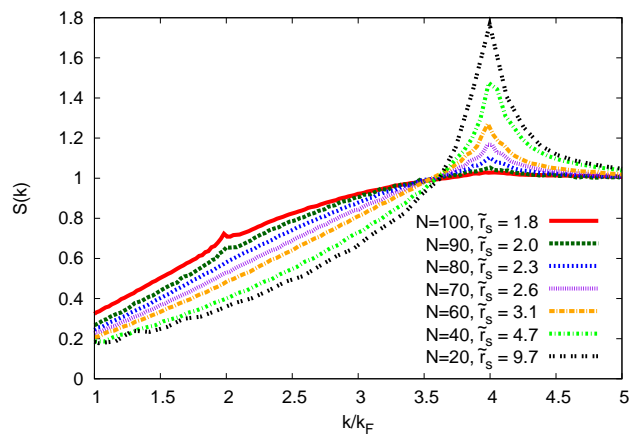}
\caption{(Color online) Static structure factor for a wire as in
Fig.~\ref{sk-wirescr}, but with finite length $S(k)$ is plotted for $20$,
$40$, $60$, $70$, $80$, $90$ and $100$ electrons. The corresponding
effective densities $\tilde{r}_s$ are reported in the legend.} 
\label{sk-finite}
\end{figure}

Again the static structure factor is determined for different densities of
electrons in the upper wire.  In contrast to the calculations for the
homogeneous system, the density of the electrons is not a direct input to
the calculation.  Instead, we control the number of electrons in the wire
which are then free to relax according to the external potential. An average
density can be determined by considering the locations of the $2\tilde{k}_F$
and $4\tilde{k}_F$ peaks of the structure factor and comparing their value
to those of an infinite array of electrons, $2\tilde{k}_F =
\frac{\pi}{2\tilde{r}_s}$ and $4\tilde{k}_F = \frac{\pi}{\tilde{r}_s}$,
$\tilde{r}_s$ being the effective density in the system. Using these
conventions, the structure factor for several different numbers of electrons
is plotted in Fig.~\ref{sk-finite}.

\begin{figure}[!ht]
\centering
\includegraphics[angle=-90,width=\columnwidth]{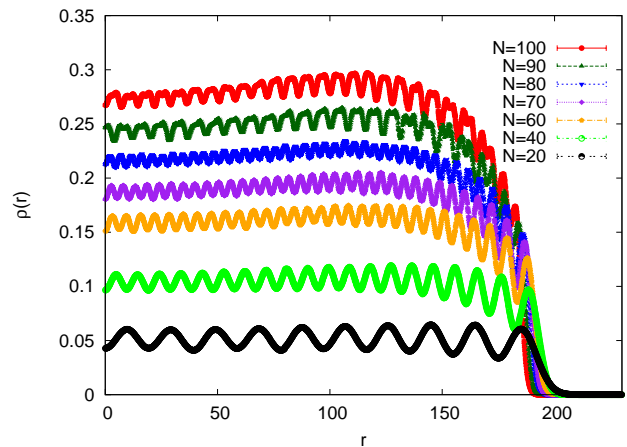}
\caption{(Color online) Density profile for electrons in the finite wire as
in Fig.~\ref{sk-finite}, plotted for half of the wire length. $N = 20$,
$40$, $60$, $70$, $80$, $90$ and $100$ are considered.} 
\label{rho-finite}
\end{figure}

In addition to the formation of a broad peak in the $S(k)$ at $4 k_F$ around
$N=80$, which corresponds to $\tilde{r}_s = 2.3$, the density profile
$n(x)=\langle \sum_i\delta(x-x_i) \rangle$ of the electrons also shows a
clear cut sign of the transition. At low densities, the electrons are
distributed in order to minimize the interparticle repulsion. This leads to
$N$ oscillations in the density profile of the wire, a configuration also
called ``Wigner molecule'',\cite{hausler-wigner-molecule} which corresponds
to the $4k_F$ peak in the $S(k)$.  When the density is increased, the number
of peaks in the density profile is reduced by a factor of two, the Pauli
exclusion principle between like spin particles being the only factor that
prevents the electrons form crossing each other.  At the same time the
$4k_F$ peak in the $S(k)$ disappears and only a $2 k_F$ singularity is
present. The density is plotted in Fig.~\ref{rho-finite} for half
of the wire as the system is symmetric under inversion around its center.
This plot also suggests a transition near $N = 80$.

Surprisingly, the calculations with the confinement potential and the infinite wire
give very similar structure factors in the vicinity of the transition,
suggesting that the interparticle correlations are not strongly affected by
the external confinement at those densities.  At lower densities the peak
at $4k_F$ is much larger for the homogeneous system because of the limited
number of particles in the finite wire.  Both the infinite and finite wires
show a transition from a system with $2k_F$ correlations to a state where
correlations have a $2 r_s$ periodicity. The crossover occurs around $r_s = 2.3$, 
which corresponds to the density of $22\mu m^{-1}$ in a GaAs
heterostructure. This is very close to the density found by Steinberg \emph{et al.}
$(20\mu m^{-1})$ for the localization transition in wires where one subband is occupied.
However it seems that in the experiment the localization involves only few
particles (up to $12$ in the highest density localized state), i.e. only a
section of the wire takes part in the transition.  This is an important
difference with respect to our calculations where the transition takes
place throughout the system in a quite homogeneous way. A non homogeneous
behavior is found at the edge of the wire where the confining potential in
Eq.~\ref{confinement} turns upward. There the transition happens at higher
densities, as one can see in Fig.~\ref{rho-finite}. This can be understood
in terms of a local mean field description. At the edge of the wire the effective
chemical potential $\mu_0 - V(x)$ is smaller, corresponding locally to a
fluid at much lower density.  

Apart from these features, we did not find any Wigner correlated patch
embedded in a liquid-like system, which seems to be the experimental
outcome. Therefore a more detailed analysis of the experimental setup is
required to understand better the experiment. For instance, one of the top
metallic gates used to tune the upper wire density could induce a plateau in
the external potential, nucleating a Wigner region as suggested by
Mueller.\cite{mueller}  On the other hand, the role of disorder is not
clear.  Although in the liquid phase the system is in a ballistic regime,
when the conductance is quantized the disorder could take over in the
localized phase and affect the charge distribution in the wire.  AlAs
wires, where the disorder is stronger, revealed conductance resonances
explained in terms of Coulomb blockade (CB) physics.\cite{grayson-alas} 
CB behavior has also been found in the localized phase of GaAs
wires.\cite{steinberg-tunneling-depleted-top-wire} 

Even if there are features that still need explanation, our calculations
show that the electronic correlation plays a very important role at the
experimental conditions, as the $2k_F$-to-$4k_F$ correlations transition
takes place exactly in the proximity of the critical density for
localization found in the experiment.  In addition to this result, which is the main
outcome of the paper, we also determined the charge and spin velocities by
means of the QMC method explained in Sec.~\ref{unscreened_coulomb_interactions} 
and the effective $J$ coupling via the WKB approach. We computed those quantities
close to the transition for the homogeneous wire with $r_s=1.25$ ($40 \mu m^{-1}$).
The charge velocity turns out to be $v_\rho= 2.33 v_F$. The
corresponding LL parameter $g=v_F/v_\rho=0.43$ is in agreement with previous
estimates\cite{fiete} and comparable to the experimental $g \approx 0.5$, measured at the
density of $40 \mu m^{-1}$.\cite{auslaender-science1}  At the same density the experimental value for
$v_F/v_\sigma$ is in the range of $1.1-1.6$, while we found
$v_F/v_\sigma=1.24$ at $40 \mu m^{-1}$.  

\begin{figure}[!ht]
\centering
\includegraphics[angle=-90,width=\columnwidth]{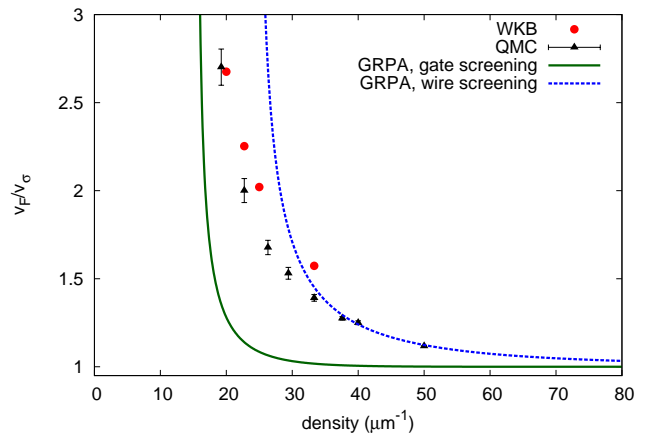}
\caption{(Color online)  Inverse spin velocity of the infinite wire.  The
red circles indicate estimates from the WKB approximation, whereas the black
triangles are determined using the QMC method described in Section
\ref{unscreened_coulomb_interactions}.  The two lines are estimates due to
the perturbative generalized random phase approximation (GRPA)\cite{GRPA}:
$v_F/v_\sigma^{GRPA} = 1/\sqrt{1 - V(2k_F)/(\pi v_F)}$.  The green line for
the gated wire (with $R=50$) uses the potential described by Auslaender \emph{et
al.}\cite{auslaender-science1} whereas the dotted blue line uses the
potential (Eq.~\ref{wire_screened_interaction}) screened by 
the lower wire.} 
\label{vsigma-wires}
\end{figure}

In Fig.~\ref{vsigma-wires} we plot the full dependence of 
the spin velocities on the density computed with the perturbative generalized RPA
(GRPA),\cite{GRPA} WKB and the exact QMC methods.  Although the GRPA is
poor near the localization transition, it agrees with the QMC at high density.
As noted above the experimentally measured spin velocities are also in 
rough agreement with the QMC estimate in a range of densities around $n = 40
\mu m^{-1}$.  To show the importance of the microscopic details of the
interaction in reproducing the measured values we also display in Fig.~\ref{vsigma-wires} 
the GRPA prediction based on a different model potential which assumes a screening due to a metallic gate
at $R = 50$.\cite{auslaender-science1}  This latter model gives virtually
unrenormalized spin velocities ($v_\sigma
\approx v_F$) up to $n=40 \mu m^{-1}$ in contrast with the strong 
suppression of the values found in the experiment.

Last but not least, our WKB estimate of $J$ turns out to be of the order of the
experimental temperature ($T=0.25 K$) around $n = 10 \mu m^{-1}$.
This means that at least the first few Coulomb blockade peaks in the
experiment should be in a spin incoherent regime, where the LL description
by Fiete \emph{et al.} applies, although in the vicinity of the transition
the spin degrees of freedom are not dominated by thermal broadening.

\section{Conclusions}
\label{conclusions}
We have presented extensive quantum Monte Carlo calculations to study the
properties of electrons constrained to one dimension with a harmonic confinement 
and interacting via several different potentials.

For unscreened interactions
with a long range $1/x$ tail there are three different regimes.
At high density the electrons behave as a
correlated liquid, transitioning to a quasi Wigner crystal 
as the density decreases,
where strong $4 k_F$ correlations follow the LL predictions.\cite{schulz} 
We accurately determined the crossover density for various thicknesses 
and found that the crossover is pushed 
to higher densities for thinner wires.
Finally at very low densities the charge degrees
of freedom are described by spinless fermions and the spins decouple 
with exponentially small exchange interactions.
We approached this limit by using
the WKB approximation.

When screening is introduced, the interactions are not long-range, and the
quasi Wigner crystal order is destroyed. However, $4 k_F$
correlations are still present even in the case of screened
interactions.  The spinless fermion regime acquires a new behavior when
the wire is very thin and the screening makes the potential
short-range. In this case the particles act as though they were
noninteracting and spinless in analogy to physics
previously studied for bosons with an infinite contact repulsion.\cite{girardeau}

We applied our numerical approach to analyze a model chosen to realistically describe 
the double wire system studied 
in the experiments of Steinberg \emph{et al.},\cite{steinberg-tunneling-depleted-top-wire}   
where a localization transition is observed.
Our model assumes screening due to a second wire described within linear response theory,
and includes the finite length of the wire 
via the external potential derived in Ref.~\onlinecite{soft-confinement-prb}.
We show that a crossover from a liquid to a state with $4 k_F$ correlations occurs
around the localization density found in the experiment.
Additionally, our exact Monte Carlo calculations yield charge and spin velocities 
for this model in agreement with those observed in the experiment close
to the transition.  We stress that the observables such as the
transition density and the spin velocity are particularly sensitive to the
microscopic details of the model interaction.  To reproduce all features
of the experiment it may be necessary to include further refinements such as a 
more accurate modulation of the external potential due to the gates, the
effects of higher subbands in the transverse
direction and the full treatment of interwire electronic correlation by
explicitly including the electrons in the other wire.
However, the simple model considered here shows that the exact treatment of
electronic correlation is essential to quantitatively describe the localization
transition seen in experiments.

\acknowledgments 
We thank D. M. Ceperley, S. Vishveshwara, O. Auslaender and M. Grayson for useful
discussions.  L.S., M.C., and R.M.M. acknowledge support in the form of the
NSF grant DMR-0404853.

\bibliography{wire_range_width}

\begin{thebibliography}{53}
\expandafter\ifx\csname natexlab\endcsname\relax\def\natexlab#1{#1}\fi
\expandafter\ifx\csname bibnamefont\endcsname\relax
  \def\bibnamefont#1{#1}\fi
\expandafter\ifx\csname bibfnamefont\endcsname\relax
  \def\bibfnamefont#1{#1}\fi
\expandafter\ifx\csname citenamefont\endcsname\relax
  \def\citenamefont#1{#1}\fi
\expandafter\ifx\csname url\endcsname\relax
  \def\url#1{\texttt{#1}}\fi
\expandafter\ifx\csname urlprefix\endcsname\relax\def\urlprefix{URL }\fi
\providecommand{\bibinfo}[2]{#2}
\providecommand{\eprint}[2][]{\url{#2}}

\bibitem[{\citenamefont{{A. M. Chang}}(2003)}]{chang}
\bibinfo{author}{\bibnamefont{{A. M. Chang}}}, \bibinfo{journal}{\rmp}
  \textbf{\bibinfo{volume}{75}}, \bibinfo{pages}{1449} (\bibinfo{year}{2003}).

\bibitem[{\citenamefont{{J. Voit}}(1995)}]{voit}
\bibinfo{author}{\bibnamefont{{J. Voit}}}, \bibinfo{journal}{Rep. Prog. Phys.}
  \textbf{\bibinfo{volume}{58}}, \bibinfo{pages}{977} (\bibinfo{year}{1995}).

\bibitem[{\citenamefont{Giamarchi}(2004)}]{giamarchi-book}
\bibinfo{author}{\bibfnamefont{T.}~\bibnamefont{Giamarchi}},
  \emph{\bibinfo{title}{Quantum Physics in One Dimension}}
  (\bibinfo{publisher}{Clarendon Press}, \bibinfo{address}{Oxford},
  \bibinfo{year}{2004}).

\bibitem[{\citenamefont{{J. M. Luttinger}}(1963)}]{luttinger}
\bibinfo{author}{\bibnamefont{{J. M. Luttinger}}}, \bibinfo{journal}{J. Math.
  Phys.} \textbf{\bibinfo{volume}{4}}, \bibinfo{pages}{1154}
  (\bibinfo{year}{1963}).

\bibitem[{\citenamefont{{S. Tomonaga}}(1963)}]{tomonaga}
\bibinfo{author}{\bibnamefont{{S. Tomonaga}}}, \bibinfo{journal}{Prog. Theor.
  Phys.} \textbf{\bibinfo{volume}{5}}, \bibinfo{pages}{554}
  (\bibinfo{year}{1963}).

\bibitem[{\citenamefont{{H. Steinberg} et~al.}(2006)\citenamefont{{H.
  Steinberg}, {O. M. Auslaender}, {A. Yacoby}, {J. Qian}, {G. A. Fiete}, {Y.
  Tserkovnyak}, {B. I. Halperin}, {K. W. Baldwin}, {L. N. Pfeiffer}, and {K. W.
  West}}}]{steinberg-tunneling-depleted-top-wire}
\bibinfo{author}{\bibnamefont{{H. Steinberg}}},
  \bibinfo{author}{\bibnamefont{{O. M. Auslaender}}},
  \bibinfo{author}{\bibnamefont{{A. Yacoby}}},
  \bibinfo{author}{\bibnamefont{{J. Qian}}}, \bibinfo{author}{\bibnamefont{{G.
  A. Fiete}}}, \bibinfo{author}{\bibnamefont{{Y. Tserkovnyak}}},
  \bibinfo{author}{\bibnamefont{{B. I. Halperin}}},
  \bibinfo{author}{\bibnamefont{{K. W. Baldwin}}},
  \bibinfo{author}{\bibnamefont{{L. N. Pfeiffer}}}, \bibnamefont{and}
  \bibinfo{author}{\bibnamefont{{K. W. West}}}, \bibinfo{journal}{\prb}
  \textbf{\bibinfo{volume}{77}}, \bibinfo{pages}{113307}
  (\bibinfo{year}{2006}).

\bibitem[{\citenamefont{{O. M. Auslaender} et~al.}(2005)\citenamefont{{O. M.
  Auslaender}, {H. Steinberg}, {A. Yacoby}, {Y. Tserkovnyak}, {B. I. Halperin},
  {K. W. Baldwin}, {L. N. Pfeiffer}, and {K. W. West}}}]{auslaender-science1}
\bibinfo{author}{\bibnamefont{{O. M. Auslaender}}},
  \bibinfo{author}{\bibnamefont{{H. Steinberg}}},
  \bibinfo{author}{\bibnamefont{{A. Yacoby}}},
  \bibinfo{author}{\bibnamefont{{Y. Tserkovnyak}}},
  \bibinfo{author}{\bibnamefont{{B. I. Halperin}}},
  \bibinfo{author}{\bibnamefont{{K. W. Baldwin}}},
  \bibinfo{author}{\bibnamefont{{L. N. Pfeiffer}}}, \bibnamefont{and}
  \bibinfo{author}{\bibnamefont{{K. W. West}}}, \bibinfo{journal}{Science}
  \textbf{\bibinfo{volume}{308}}, \bibinfo{pages}{88} (\bibinfo{year}{2005}).

\bibitem[{\citenamefont{{Y. Tserkovnyak} et~al.}(2002)\citenamefont{{Y.
  Tserkovnyak}, {B. Halperin}, {O. Auslaender}, and {A.
  Yacoby}}}]{soft-confinement-prl}
\bibinfo{author}{\bibnamefont{{Y. Tserkovnyak}}},
  \bibinfo{author}{\bibnamefont{{B. Halperin}}},
  \bibinfo{author}{\bibnamefont{{O. Auslaender}}}, \bibnamefont{and}
  \bibinfo{author}{\bibnamefont{{A. Yacoby}}}, \bibinfo{journal}{\prl}
  \textbf{\bibinfo{volume}{89}}, \bibinfo{pages}{136805}
  (\bibinfo{year}{2002}).

\bibitem[{\citenamefont{{Y. Tserkovnyak} et~al.}(2003)\citenamefont{{Y.
  Tserkovnyak}, {B. Halperin}, {O. Auslaender}, and {A.
  Yacoby}}}]{soft-confinement-prb}
\bibinfo{author}{\bibnamefont{{Y. Tserkovnyak}}},
  \bibinfo{author}{\bibnamefont{{B. Halperin}}},
  \bibinfo{author}{\bibnamefont{{O. Auslaender}}}, \bibnamefont{and}
  \bibinfo{author}{\bibnamefont{{A. Yacoby}}}, \bibinfo{journal}{\prb}
  \textbf{\bibinfo{volume}{68}}, \bibinfo{pages}{125312}
  (\bibinfo{year}{2003}).

\bibitem[{\citenamefont{{O. M. Auslaender} et~al.}(2002)\citenamefont{{O. M.
  Auslaender}, {A. Yacoby}, {R. de Picciotto}, {K. W. Baldwin}, {L. N.
  Pfeiffer}, and {K. W. West}}}]{auslaender-science0}
\bibinfo{author}{\bibnamefont{{O. M. Auslaender}}},
  \bibinfo{author}{\bibnamefont{{A. Yacoby}}},
  \bibinfo{author}{\bibnamefont{{R. de Picciotto}}},
  \bibinfo{author}{\bibnamefont{{K. W. Baldwin}}},
  \bibinfo{author}{\bibnamefont{{L. N. Pfeiffer}}}, \bibnamefont{and}
  \bibinfo{author}{\bibnamefont{{K. W. West}}}, \bibinfo{journal}{Science}
  \textbf{\bibinfo{volume}{295}}, \bibinfo{pages}{825} (\bibinfo{year}{2002}).

\bibitem[{\citenamefont{{O. M. Auslaender} et~al.}(2004)\citenamefont{{O. M.
  Auslaender}, {H. Steinberg}, {A. Yacoby}, {Y. Tserkovnyak}, {B. I. Halperin},
  {R. de Picciotto}, {K. W. Baldwin}, {L. N. Pfeiffer}, and {K. W.
  West}}}]{auslaender-solid-state-comm}
\bibinfo{author}{\bibnamefont{{O. M. Auslaender}}},
  \bibinfo{author}{\bibnamefont{{H. Steinberg}}},
  \bibinfo{author}{\bibnamefont{{A. Yacoby}}},
  \bibinfo{author}{\bibnamefont{{Y. Tserkovnyak}}},
  \bibinfo{author}{\bibnamefont{{B. I. Halperin}}},
  \bibinfo{author}{\bibnamefont{{R. de Picciotto}}},
  \bibinfo{author}{\bibnamefont{{K. W. Baldwin}}},
  \bibinfo{author}{\bibnamefont{{L. N. Pfeiffer}}}, \bibnamefont{and}
  \bibinfo{author}{\bibnamefont{{K. W. West}}}, \bibinfo{journal}{Sol. Stat.
  Comm.} \textbf{\bibinfo{volume}{131}}, \bibinfo{pages}{657}
  (\bibinfo{year}{2004}).

\bibitem[{\citenamefont{{A. Yacoby} et~al.}(2006)\citenamefont{{A. Yacoby}, {O.
  M. Auslaender}, {H. Steinberg}, {Y. Tserkovnyak}, {B. I. Halperin}, {K. W.
  Baldwin}, {L. N. Pfeiffer}, and {K. W. West}}}]{yacoby}
\bibinfo{author}{\bibnamefont{{A. Yacoby}}}, \bibinfo{author}{\bibnamefont{{O.
  M. Auslaender}}}, \bibinfo{author}{\bibnamefont{{H. Steinberg}}},
  \bibinfo{author}{\bibnamefont{{Y. Tserkovnyak}}},
  \bibinfo{author}{\bibnamefont{{B. I. Halperin}}},
  \bibinfo{author}{\bibnamefont{{K. W. Baldwin}}},
  \bibinfo{author}{\bibnamefont{{L. N. Pfeiffer}}}, \bibnamefont{and}
  \bibinfo{author}{\bibnamefont{{K. W. West}}}, \bibinfo{journal}{Phys. Stat.
  Sol. B} \textbf{\bibinfo{volume}{243}}, \bibinfo{pages}{3593}
  (\bibinfo{year}{2006}).

\bibitem[{\citenamefont{Foulkes et~al.}(2001)\citenamefont{Foulkes, Mitas,
  Needs, and Rajagopal}}]{RevModPhys_foulkes}
\bibinfo{author}{\bibfnamefont{W.~M.~C.} \bibnamefont{Foulkes}},
  \bibinfo{author}{\bibfnamefont{L.}~\bibnamefont{Mitas}},
  \bibinfo{author}{\bibfnamefont{R.~J.} \bibnamefont{Needs}}, \bibnamefont{and}
  \bibinfo{author}{\bibfnamefont{G.}~\bibnamefont{Rajagopal}},
  \bibinfo{journal}{Rev. Mod. Phys.} \textbf{\bibinfo{volume}{73}},
  \bibinfo{pages}{33} (\bibinfo{year}{2001}).

\bibitem[{\citenamefont{Reynolds et~al.}(1982)\citenamefont{Reynolds, Ceperley,
  Alder, and Lester}}]{reynolds_dmc}
\bibinfo{author}{\bibfnamefont{P.~J.} \bibnamefont{Reynolds}},
  \bibinfo{author}{\bibfnamefont{D.~M.} \bibnamefont{Ceperley}},
  \bibinfo{author}{\bibfnamefont{B.~J.} \bibnamefont{Alder}}, \bibnamefont{and}
  \bibinfo{author}{\bibfnamefont{W.~A.} \bibnamefont{Lester}},
  \bibinfo{journal}{The Journal of Chemical Physics}
  \textbf{\bibinfo{volume}{77}}, \bibinfo{pages}{5593} (\bibinfo{year}{1982}),
  \urlprefix\url{http://link.aip.org/link/?JCP/77/5593/1}.

\bibitem[{\citenamefont{Umrigar et~al.}(1993)\citenamefont{Umrigar,
  Nightingale, and Runge}}]{umrigar_dmc}
\bibinfo{author}{\bibfnamefont{C.~J.} \bibnamefont{Umrigar}},
  \bibinfo{author}{\bibfnamefont{M.~P.} \bibnamefont{Nightingale}},
  \bibnamefont{and} \bibinfo{author}{\bibfnamefont{K.~J.} \bibnamefont{Runge}},
  \bibinfo{journal}{The Journal of Chemical Physics}
  \textbf{\bibinfo{volume}{99}}, \bibinfo{pages}{2865} (\bibinfo{year}{1993}),
  \urlprefix\url{http://link.aip.org/link/?JCP/99/2865/1}.

\bibitem[{\citenamefont{Casula et~al.}(2005)\citenamefont{Casula, Filippi, and
  Sorella}}]{casula_lrdmc}
\bibinfo{author}{\bibfnamefont{M.}~\bibnamefont{Casula}},
  \bibinfo{author}{\bibfnamefont{C.}~\bibnamefont{Filippi}}, \bibnamefont{and}
  \bibinfo{author}{\bibfnamefont{S.}~\bibnamefont{Sorella}},
  \bibinfo{journal}{Physical Review Letters} \textbf{\bibinfo{volume}{95}},
  \bibinfo{eid}{100201} (pages~\bibinfo{numpages}{4}) (\bibinfo{year}{2005}),
  \urlprefix\url{http://link.aps.org/abstract/PRL/v95/e100201}.

\bibitem[{\citenamefont{{W. H\"ausler} et~al.}(2002)\citenamefont{{W.
  H\"ausler}, {L. Kecke}, and {A. H. MacDonald}}}]{hausler-qmc-quantum-wires}
\bibinfo{author}{\bibnamefont{{W. H\"ausler}}},
  \bibinfo{author}{\bibnamefont{{L. Kecke}}}, \bibnamefont{and}
  \bibinfo{author}{\bibnamefont{{A. H. MacDonald}}}, \bibinfo{journal}{Phys.
  Rev. B} \textbf{\bibinfo{volume}{65}}, \bibinfo{pages}{085104}
  (\bibinfo{year}{2002}).

\bibitem[{\citenamefont{Casula et~al.}(2006)\citenamefont{Casula, {S. Sorella},
  and {G. Senatore}}}]{casula-1d}
\bibinfo{author}{\bibfnamefont{M.}~\bibnamefont{Casula}},
  \bibinfo{author}{\bibnamefont{{S. Sorella}}}, \bibnamefont{and}
  \bibinfo{author}{\bibnamefont{{G. Senatore}}}, \bibinfo{journal}{\prb}
  \textbf{\bibinfo{volume}{74}}, \bibinfo{pages}{245427}
  (\bibinfo{year}{2006}).

\bibitem[{\citenamefont{Wigner}(1934)}]{wigner-crystal}
\bibinfo{author}{\bibfnamefont{E.~P.} \bibnamefont{Wigner}},
  \bibinfo{journal}{Phys. Rev.} \textbf{\bibinfo{volume}{46}},
  \bibinfo{pages}{1002} (\bibinfo{year}{1934}).

\bibitem[{\citenamefont{{P. Nozi\`eres} and {D. Pines}}(1999)}]{pines-book}
\bibinfo{author}{\bibnamefont{{P. Nozi\`eres}}} \bibnamefont{and}
  \bibinfo{author}{\bibnamefont{{D. Pines}}}, \emph{\bibinfo{title}{The Theory
  of Quantum Liquids}} (\bibinfo{publisher}{Perseus},
  \bibinfo{address}{Cambridge, MA}, \bibinfo{year}{1999}),
  \bibinfo{edition}{3rd} ed.

\bibitem[{\citenamefont{{D. M. Ceperley} and {B. J.
  Alder}}(1980)}]{ceperley-alder}
\bibinfo{author}{\bibnamefont{{D. M. Ceperley}}} \bibnamefont{and}
  \bibinfo{author}{\bibnamefont{{B. J. Alder}}}, \bibinfo{journal}{\prl}
  \textbf{\bibinfo{volume}{45}}, \bibinfo{pages}{566} (\bibinfo{year}{1980}).

\bibitem[{\citenamefont{{B. Bernu} et~al.}(2001)\citenamefont{{B. Bernu}, {D.
  M. Ceperley}, and {L. Candido}}}]{ceperley-exchange}
\bibinfo{author}{\bibnamefont{{B. Bernu}}}, \bibinfo{author}{\bibnamefont{{D.
  M. Ceperley}}}, \bibnamefont{and} \bibinfo{author}{\bibnamefont{{L.
  Candido}}}, \bibinfo{journal}{\prl} \textbf{\bibinfo{volume}{86}},
  \bibinfo{pages}{870} (\bibinfo{year}{2001}).

\bibitem[{\citenamefont{{H. J. Schulz}}(1993)}]{schulz}
\bibinfo{author}{\bibnamefont{{H. J. Schulz}}}, \bibinfo{journal}{\prl}
  \textbf{\bibinfo{volume}{71}}, \bibinfo{pages}{1864} (\bibinfo{year}{1993}).

\bibitem[{\citenamefont{{M.
  Fogler}}(2005{\natexlab{a}})}]{fogler-coulomb-tonks-prl}
\bibinfo{author}{\bibnamefont{{M. Fogler}}}, \bibinfo{journal}{\prl}
  \textbf{\bibinfo{volume}{94}}, \bibinfo{pages}{056405}
  (\bibinfo{year}{2005}{\natexlab{a}}).

\bibitem[{\citenamefont{{M.
  Fogler}}(2005{\natexlab{b}})}]{fogler-coulomb-tonks-prb}
\bibinfo{author}{\bibnamefont{{M. Fogler}}}, \bibinfo{journal}{\prb}
  \textbf{\bibinfo{volume}{71}}, \bibinfo{pages}{161304}
  (\bibinfo{year}{2005}{\natexlab{b}}).

\bibitem[{\citenamefont{Girardeau}(1960)}]{girardeau}
\bibinfo{author}{\bibfnamefont{M.}~\bibnamefont{Girardeau}},
  \bibinfo{journal}{J. Math. Phys.} \textbf{\bibinfo{volume}{1}},
  \bibinfo{pages}{516} (\bibinfo{year}{1960}).

\bibitem[{\citenamefont{{W. Hauseler} and {RB. Kramer}}(1996)}]{lanczos1}
\bibinfo{author}{\bibnamefont{{W. Hauseler}}} \bibnamefont{and}
  \bibinfo{author}{\bibnamefont{{RB. Kramer}}}, \bibinfo{journal}{\prb}
  \textbf{\bibinfo{volume}{47}}, \bibinfo{pages}{16353} (\bibinfo{year}{1996}).

\bibitem[{\citenamefont{{K. Jauregui} et~al.}(1993)\citenamefont{{K. Jauregui},
  {W. Hauseler}, and {B. Kramer}}}]{lanczos2}
\bibinfo{author}{\bibnamefont{{K. Jauregui}}},
  \bibinfo{author}{\bibnamefont{{W. Hauseler}}}, \bibnamefont{and}
  \bibinfo{author}{\bibnamefont{{B. Kramer}}}, \bibinfo{journal}{Europhys.
  Lett.} \textbf{\bibinfo{volume}{24}}, \bibinfo{pages}{581}
  (\bibinfo{year}{1993}).

\bibitem[{\citenamefont{{K. Jauregui} et~al.}(1996)\citenamefont{{K. Jauregui},
  {W. Hauseler}, {D. Weinmann}, and {B. Kramer}}}]{lanczos3}
\bibinfo{author}{\bibnamefont{{K. Jauregui}}},
  \bibinfo{author}{\bibnamefont{{W. Hauseler}}},
  \bibinfo{author}{\bibnamefont{{D. Weinmann}}}, \bibnamefont{and}
  \bibinfo{author}{\bibnamefont{{B. Kramer}}}, \bibinfo{journal}{\prb}
  \textbf{\bibinfo{volume}{53}}, \bibinfo{pages}{R1713} (\bibinfo{year}{1996}).

\bibitem[{\citenamefont{Mueller}(2005)}]{mueller}
\bibinfo{author}{\bibfnamefont{E.~J.} \bibnamefont{Mueller}},
  \bibinfo{journal}{\prb} \textbf{\bibinfo{volume}{72}},
  \bibinfo{pages}{075322} (\bibinfo{year}{2005}).

\bibitem[{\citenamefont{{G. A. Fiete} et~al.}(2005)\citenamefont{{G. A. Fiete},
  {J. Qian}, {Y. Tserkovnyak}, and {B. I. Halperin}}}]{fiete}
\bibinfo{author}{\bibnamefont{{G. A. Fiete}}},
  \bibinfo{author}{\bibnamefont{{J. Qian}}}, \bibinfo{author}{\bibnamefont{{Y.
  Tserkovnyak}}}, \bibnamefont{and} \bibinfo{author}{\bibnamefont{{B. I.
  Halperin}}}, \bibinfo{journal}{\prb} \textbf{\bibinfo{volume}{72}},
  \bibinfo{pages}{045315} (\bibinfo{year}{2005}).

\bibitem[{\citenamefont{{A. Javey} et~al.}(2002)\citenamefont{{A. Javey}, {H.
  Kim}, {M. Brink}, {Q. Wang}, {A. Ural}, {J. Gu}, {P. McIntyre}, {P. McEuen},
  {M. Lundstrom}, and {H. Dai}}}]{ultrathin-nanotube}
\bibinfo{author}{\bibnamefont{{A. Javey}}}, \bibinfo{author}{\bibnamefont{{H.
  Kim}}}, \bibinfo{author}{\bibnamefont{{M. Brink}}},
  \bibinfo{author}{\bibnamefont{{Q. Wang}}}, \bibinfo{author}{\bibnamefont{{A.
  Ural}}}, \bibinfo{author}{\bibnamefont{{J. Gu}}},
  \bibinfo{author}{\bibnamefont{{P. McIntyre}}},
  \bibinfo{author}{\bibnamefont{{P. McEuen}}},
  \bibinfo{author}{\bibnamefont{{M. Lundstrom}}}, \bibnamefont{and}
  \bibinfo{author}{\bibnamefont{{H. Dai}}}, \bibinfo{journal}{Nat. Mater.}
  \textbf{\bibinfo{volume}{1}}, \bibinfo{pages}{241} (\bibinfo{year}{2002}).

\bibitem[{\citenamefont{{B. M. Kim} et~al.}(2004)\citenamefont{{B. M. Kim}, {T.
  Brintlinger}, {E. Cobas}, {M. S. Fuhrer}, {H. Zheng}, {Z. Yu}, {R. Droopad},
  {J. Ramdani}, and {K. Eisenbeiser}}}]{ultrathin-nanotube2}
\bibinfo{author}{\bibnamefont{{B. M. Kim}}}, \bibinfo{author}{\bibnamefont{{T.
  Brintlinger}}}, \bibinfo{author}{\bibnamefont{{E. Cobas}}},
  \bibinfo{author}{\bibnamefont{{M. S. Fuhrer}}},
  \bibinfo{author}{\bibnamefont{{H. Zheng}}}, \bibinfo{author}{\bibnamefont{{Z.
  Yu}}}, \bibinfo{author}{\bibnamefont{{R. Droopad}}},
  \bibinfo{author}{\bibnamefont{{J. Ramdani}}}, \bibnamefont{and}
  \bibinfo{author}{\bibnamefont{{K. Eisenbeiser}}}, \bibinfo{journal}{Appl.
  Phys. Lett.} \textbf{\bibinfo{volume}{84}}, \bibinfo{pages}{1946}
  (\bibinfo{year}{2004}).

\bibitem[{\citenamefont{{K. X. Liu} et~al.}(2001)\citenamefont{{K. X. Liu}, {M.
  H. Kalos}, and {G. V. Chester}}}]{forward-walking}
\bibinfo{author}{\bibnamefont{{K. X. Liu}}}, \bibinfo{author}{\bibnamefont{{M.
  H. Kalos}}}, \bibnamefont{and} \bibinfo{author}{\bibnamefont{{G. V.
  Chester}}}, \bibinfo{journal}{\prl} \textbf{\bibinfo{volume}{86}},
  \bibinfo{pages}{870} (\bibinfo{year}{2001}).

\bibitem[{\citenamefont{Calandra~Buonaura and Sorella}(1998)}]{sorellaFW}
\bibinfo{author}{\bibfnamefont{M.}~\bibnamefont{Calandra~Buonaura}}
  \bibnamefont{and} \bibinfo{author}{\bibfnamefont{S.}~\bibnamefont{Sorella}},
  \bibinfo{journal}{Phys. Rev. B} \textbf{\bibinfo{volume}{57}},
  \bibinfo{pages}{11446} (\bibinfo{year}{1998}).

\bibitem[{\citenamefont{{T. Gaskell}}(1961)}]{gaskell}
\bibinfo{author}{\bibnamefont{{T. Gaskell}}}, \bibinfo{journal}{Proc. Phys.
  Soc.} \textbf{\bibinfo{volume}{77}}, \bibinfo{pages}{1182}
  (\bibinfo{year}{1961}).

\bibitem[{\citenamefont{{A. Gold} and {L. Calmels}}(1995)}]{gold-rpa-msa}
\bibinfo{author}{\bibnamefont{{A. Gold}}} \bibnamefont{and}
  \bibinfo{author}{\bibnamefont{{L. Calmels}}}, \bibinfo{journal}{Solid State
  Comm.} \textbf{\bibinfo{volume}{96}}, \bibinfo{pages}{101}
  (\bibinfo{year}{1995}).

\bibitem[{\citenamefont{Casula and {G.
  Senatore}}(2005)}]{casula-1d-correlations}
\bibinfo{author}{\bibfnamefont{M.}~\bibnamefont{Casula}} \bibnamefont{and}
  \bibinfo{author}{\bibnamefont{{G. Senatore}}},
  \bibinfo{journal}{ChemPhysChem} \textbf{\bibinfo{volume}{6}},
  \bibinfo{pages}{1902} (\bibinfo{year}{2005}).

\bibitem[{\citenamefont{Giamarchi and Schulz}(1989)}]{giamarchi}
\bibinfo{author}{\bibfnamefont{T.}~\bibnamefont{Giamarchi}} \bibnamefont{and}
  \bibinfo{author}{\bibfnamefont{H.~J.} \bibnamefont{Schulz}},
  \bibinfo{journal}{Phys. Rev. B} \textbf{\bibinfo{volume}{39}},
  \bibinfo{pages}{4620} (\bibinfo{year}{1989}).

\bibitem[{\citenamefont{{L. Shulenburger} et~al.}()\citenamefont{{L.
  Shulenburger}, {M. Casula}, {G. Senatore}, and {R. M.
  Martin}}}]{shulenburger}
\bibinfo{author}{\bibnamefont{{L. Shulenburger}}},
  \bibinfo{author}{\bibnamefont{{M. Casula}}},
  \bibinfo{author}{\bibnamefont{{G. Senatore}}}, \bibnamefont{and}
  \bibinfo{author}{\bibnamefont{{R. M. Martin}}}, \bibinfo{note}{{\em in
  preparation}}.

\bibitem[{\citenamefont{{A. Gold} and {L. Calmels}}(1998)}]{gold-excit}
\bibinfo{author}{\bibnamefont{{A. Gold}}} \bibnamefont{and}
  \bibinfo{author}{\bibnamefont{{L. Calmels}}}, \bibinfo{journal}{\prb}
  \textbf{\bibinfo{volume}{58}}, \bibinfo{pages}{3497} (\bibinfo{year}{1998}).

\bibitem[{\citenamefont{{D. M. Ceperley} and {B. Bernu}}(1988)}]{ceperley-cfmc}
\bibinfo{author}{\bibnamefont{{D. M. Ceperley}}} \bibnamefont{and}
  \bibinfo{author}{\bibnamefont{{B. Bernu}}}, \bibinfo{journal}{\jcp}
  \textbf{\bibinfo{volume}{89}}, \bibinfo{pages}{6316} (\bibinfo{year}{1988}).

\bibitem[{\citenamefont{{R. P. Feynman} and {M.
  Cohen}}(1956)}]{feynman-helium-excitations}
\bibinfo{author}{\bibnamefont{{R. P. Feynman}}} \bibnamefont{and}
  \bibinfo{author}{\bibnamefont{{M. Cohen}}}, \bibinfo{journal}{Phys. Rev.}
  \textbf{\bibinfo{volume}{102}}, \bibinfo{pages}{1189} (\bibinfo{year}{1956}).

\bibitem[{\citenamefont{{K. A.
  Matveev}}(2004)}]{matveev-conductance-of-quantum-wire}
\bibinfo{author}{\bibnamefont{{K. A. Matveev}}}, \bibinfo{journal}{\prb}
  \textbf{\bibinfo{volume}{70}}, \bibinfo{pages}{245319}
  (\bibinfo{year}{2004}).

\bibitem[{\citenamefont{{J. des Cloizeaux} and {J. J.
  Pearson}}(1962)}]{bethe-ansatz1}
\bibinfo{author}{\bibnamefont{{J. des Cloizeaux}}} \bibnamefont{and}
  \bibinfo{author}{\bibnamefont{{J. J. Pearson}}}, \bibinfo{journal}{Phys.
  Rev.} \textbf{\bibinfo{volume}{128}}, \bibinfo{pages}{2131}
  (\bibinfo{year}{1962}).

\bibitem[{\citenamefont{{L. D. Faddeev} and {L. A.
  Takhtajan}}(1981)}]{bethe-ansatz2}
\bibinfo{author}{\bibnamefont{{L. D. Faddeev}}} \bibnamefont{and}
  \bibinfo{author}{\bibnamefont{{L. A. Takhtajan}}}, \bibinfo{journal}{Phys.
  Lett.} \textbf{\bibinfo{volume}{85A}}, \bibinfo{pages}{375}
  (\bibinfo{year}{1981}).

\bibitem[{\citenamefont{{M. Fogler} and {E.
  Pivovarov}}(2005)}]{fogler-exchange-quantum-rings}
\bibinfo{author}{\bibnamefont{{M. Fogler}}} \bibnamefont{and}
  \bibinfo{author}{\bibnamefont{{E. Pivovarov}}}, \bibinfo{journal}{\prb}
  \textbf{\bibinfo{volume}{72}}, \bibinfo{pages}{195344}
  (\bibinfo{year}{2005}).

\bibitem[{\citenamefont{{M. Fogler} and {E.
  Pivovarov}}(2006)}]{fogler-spin-exchange}
\bibinfo{author}{\bibnamefont{{M. Fogler}}} \bibnamefont{and}
  \bibinfo{author}{\bibnamefont{{E. Pivovarov}}}, \bibinfo{journal}{J. Phys:
  Condens. Matter} \textbf{\bibinfo{volume}{18}}, \bibinfo{pages}{L7}
  (\bibinfo{year}{2006}).

\bibitem[{\citenamefont{Sorella}(2001)}]{sorellaSR}
\bibinfo{author}{\bibfnamefont{S.}~\bibnamefont{Sorella}},
  \bibinfo{journal}{Phys. Rev. B} \textbf{\bibinfo{volume}{64}},
  \bibinfo{pages}{024512} (\bibinfo{year}{2001}).

\bibitem[{\citenamefont{{Michele Casula} et~al.}(2004)\citenamefont{{Michele
  Casula}, {Claudio Attaccalite}, and {Sandro Sorella}}}]{casulaSR}
\bibinfo{author}{\bibnamefont{{Michele Casula}}},
  \bibinfo{author}{\bibnamefont{{Claudio Attaccalite}}}, \bibnamefont{and}
  \bibinfo{author}{\bibnamefont{{Sandro Sorella}}}, \bibinfo{journal}{\jcp}
  \textbf{\bibinfo{volume}{121}}, \bibinfo{pages}{7710} (\bibinfo{year}{2004}).

\bibitem[{\citenamefont{H\"ausler and Kramer}(1993)}]{hausler-wigner-molecule}
\bibinfo{author}{\bibfnamefont{W.}~\bibnamefont{H\"ausler}} \bibnamefont{and}
  \bibinfo{author}{\bibfnamefont{B.}~\bibnamefont{Kramer}},
  \bibinfo{journal}{Phys. Rev. B} \textbf{\bibinfo{volume}{47}},
  \bibinfo{pages}{16353} (\bibinfo{year}{1993}).

\bibitem[{\citenamefont{{J. Moser} et~al.}(2006)\citenamefont{{J. Moser}, {S.
  Roddaro}, {D. Schuh}, {M. Bichler}, {V. Pellegrini}, and {M.
  Grayson}}}]{grayson-alas}
\bibinfo{author}{\bibnamefont{{J. Moser}}}, \bibinfo{author}{\bibnamefont{{S.
  Roddaro}}}, \bibinfo{author}{\bibnamefont{{D. Schuh}}},
  \bibinfo{author}{\bibnamefont{{M. Bichler}}},
  \bibinfo{author}{\bibnamefont{{V. Pellegrini}}}, \bibnamefont{and}
  \bibinfo{author}{\bibnamefont{{M. Grayson}}}, \bibinfo{journal}{\prb}
  \textbf{\bibinfo{volume}{74}}, \bibinfo{pages}{193307}
  (\bibinfo{year}{2006}).

\bibitem[{\citenamefont{{C. E. Creffield} et~al.}(2001)\citenamefont{{C. E.
  Creffield}, {W. H\"ausler}, and {A. H. MacDonald}}}]{GRPA}
\bibinfo{author}{\bibnamefont{{C. E. Creffield}}},
  \bibinfo{author}{\bibnamefont{{W. H\"ausler}}}, \bibnamefont{and}
  \bibinfo{author}{\bibnamefont{{A. H. MacDonald}}},
  \bibinfo{journal}{Europhys. Lett.} \textbf{\bibinfo{volume}{53}},
  \bibinfo{pages}{221} (\bibinfo{year}{2001}).

\end{thebibliography}

\end{document}